\documentclass[12pt]{aastex}\usepackage{emulateapj5} %AASPP

\def\to{$-$}
\def\dash{$-$}
\def\HI{\protect\ion{H}{1}}
\def\CaII{\protect\ion{Ca}{2}}
\def\dex#1{10$^{#1}$}
\def\tdex#1{$\times$10$^{#1}$}
\def\deg{$^\circ$}
\def\cmm#1{cm$^{-#1}$}
\def\kms{km\,s$^{-1}$}
\def\vlsr{v$_{\rm LSR}$}

\def\Seff{2}
\def\Sother{3}
\def\Sfit{4}
\def\Sresults{5}
\def\Sanal{6}   \def\SanalHVC{6.2}

\def\Cdflag{18}

\begin{document}

\title{\HI\ spectra and column densities toward HVC and IVC probes}
\author{B.P. Wakker}
       \affil{Department of Astronomy, University of Wisconsin \\
       475 N Charter St, Madison, WI\,53706, USA \\ wakker@astro.wisc.edu}
\author{P.M.W. Kalberla}
       \affil{Radio-astronomisches Institut Universit\"at Bonn\\ D-53121 Bonn,
       Germany\\ kalberla@astro.uni-bonn.de}
\author{H. van Woerden}
       \affil{Rijks Universiteit Groningen \\ Postbus 800, 9700 AV, Groningen,
       The Netherlands \\ hugo@astro.rug.nl}
\author{K.S. de Boer}
       \affil{Sternwarte Universit\"at Bonn\\ D-53121 Bonn, Germany
       \\ deboer@astro.uni-bonn.de}
\author{M.E. Putman}
       \affil{Research School of Astronomy \& Astrophysics, Australian National
       University\\ Weston Creek P.O., Weston, ACT 2611 Australia
       \\ mary.putman@atnf.csiro.au}

\begin{abstract}
We show 21-cm line profiles in the direction of stars and extragalactic objects,
lying projected on high- and intermediate-velocity clouds (HVCs and IVCs). About
half of these are from new data obtained with the Effelsberg 100-m telescope,
about a quarter are extracted from the Leiden-Dwingeloo Survey (LDS) and the
remaining quarter were observed with other single-dish telescopes. \HI\ column
densities were determined for each HVC/IVC. Wakker (2001) (Paper~I) uses these
in combination with optical and ultraviolet high-resolution measurements to
derive abundances. Here, an analysis is given of the difference and ratio of
N(\HI) as observed with a 9\arcmin\ versus a 35\arcmin\ beam. For HVCs and IVCs
the ratio N(\HI-9\arcmin)/N(\HI-35\arcmin) lies in the range 0.2\to2.5. For
low-velocity gas this ratio ranges from 0.75 to 1.3 (the observed ratio is
0.85\to1.4, but it appears that the correction for stray radiation is slightly
off). The smaller range for the low-velocity gas may be caused by confusion in
the line of sight, so that a low ratio in one component can be compensated by a
high ratio in another \dash\ for 11 low-velocity clouds fit by one component the
distribution of ratios has a larger dispersion. Comparison with higher angular
resolution data is possible for sixteen sightlines. Eight sightlines with \HI\
data at 1\arcmin\to2\arcmin\ resolution show a range of 0.75\to1.25 for
N(\HI-2\arcmin)/N(\HI-9\arcmin), while in eight other sightlines
N(\HI-Ly$\alpha$)/N(\HI-9\arcmin) ranges from 0.74 to 0.98.
\end{abstract}


\newpage

\section{Introduction}
A description of progress in understanding distances and metallicities of high-
and intermediate-velocity neutral hydrogen (HVCs and IVCs) is given by Wakker
(2001), who lists all published data concerning ionic absorption lines in such
clouds. This paper is referred to below as ``Paper~I''. The present paper offers
a (re)analysis of 21-cm \HI\ data that were used to derive ion/\HI\ ratios. The
\HI\ data are important for two reasons: a) if absorption lines due to the HVC
or IVC are present, they allow one to derive an ion abundance for the cloud and
b) if such absorption is absent, the \HI\ data are required to derive the
expected optical line strength, in order to interpret the absence of absorption
as a lower distance limit or showing the need for deeper integrations. This is
explained more fully in Paper~I (see description of Col.~\Cdflag\ in that
paper).
\par Previous experience has suggested that a measurement with the smallest
possible beam is required. A few HVC cores have been observed at high angular
resolution (2 arcmin or better; Schwarz \& Oort 1981, Wakker \& Schwarz 1991,
Wakker et al.\ 1996, Braun \& Burton 2000). From these studies it is clear that
these cores contain much fine structure, with column density contrasts up to a
factor 5 at arcminute scales. Outside cores, however, the contrast seems to be
less (Wakker et al.\ 1996, Schwarz \& Wakker 2001). Also, the factor 5
represents the extreme contrast. For any randomly chosen direction within a
field, the contrast is closer to a factor 2 over arcminute scales.
\par It is not known whether similar contrast would occur if observations with
even smaller beams could be made. A possible way of studying this is to compare
N(\HI) as observed at 21-cm with N(\HI) derived from Ly$\alpha$ absorption
toward extra-galactic background sources. A preliminary analysis of this kind by
Savage et al.\ (2000) suggests that N(\HI-Ly$\alpha$) usually lies in the range
0.6\to1 times N(\HI-21-cm). However, with this method it is not possible to
separately measure N(\HI) for different components. For further discussion see
Sect.~\Sanal.
\par Interferometer maps (1\to2 arcmin beam) clearly are necessary to get
accurate abundances for important background probes, but it is impractical to
observe every probe that way. The next-best estimate requires a large
single-dish telescope. For probes with declinations between about 0 and +38
degrees, Arecibo (3 arcmin beam) is preferable, though data are rarely
available. We used the 100-m Effelsberg telescope (9 arcmin beam) for probes
with declinations above about $-$30 degrees. In the future the 100-m GBT (Green
Bank Telescope) may also prove useful.
\par Effelsberg data were obtained for about half of the 269 probes listed in
Paper~I (excluding the 52 stars in the LMC). For about a quarter of these
probes, other good single-dish data exist, obtained with the telescopes at
Arecibo (3 arcmin beam), Jodrell Bank (12 arcmin beam), Parkes (15 arcmin beam),
Green Bank (21 arcmin beam) and Hat Creek (36 arcmin beam, but 0.0015\,K rms).
For the remaining quarter we needed to use the Leiden-Dwingeloo Survey (LDS) of
Hartmann \& Burton (1997), which has a 35 arcmin beam. The Effelsberg data allow
us for the first time to systematically analyze the difference of N(\HI)
observed with a 9\arcmin\ beam vs a 35\arcmin\ beam, as LDS data are available
for every direction north of declination $-$35\deg.
%Half: (113+21) Effelsberg or better   134
%Quarter : good SD (6+1+17+3+1+13+14)   55
%Quarter: LDS (67)                      67
\par Section~\Seff\ gives a description of the Effelsberg observations, which
have not been presented elsewhere. Section~\Sother\ summarizes the
characteristics of the other data. Section~\Sfit\ describes the procedure used
to derive N(\HI) for the HVC and IVC components, and Sect.~\Sresults\ describes
the spectra shown in Fig.~1. That figure presents an \HI\ spectrum for each of
the probes for which absorption line measurements are listed in Paper~I.
Finally, in Sect.~\Sanal\ we make a preliminary analysis of the difference
between N(\HI) as observed at Effelsberg vs N(\HI) in the same direction as
extracted from the LDS.

\section{Effelsberg Observations}
\subsection{Summary of runs}
Observations were done on 19 dates: 12 Jan 1995, 24 Jan 1995, 26 Aug 1995, 10
Sep 1995, 11\to12 Aug 1996, 13\to14 Sep 1996, 27 Jun 1999, 12\to14 Jul 1999, 3
Mar 2000, 25\to26 Mar 2000, 3\to4 Apr 2000, 3\to4 May 2000. The large gap
between 1996 and 1999 is caused by a problem with interference that is most
probably due to Digital Audio Broadcasting (DAB).
\par The DAB transmission at a frequency near 1452 MHz was variable in intensity
and strongly polarized. The nearest transmitter at a distance of 30 km caused
interference which was received through the backlobes of the telescope, even
under the most favorable circumstances. This interference caused spurious lines
and partly a saturation of the receiver, resulting in strongly variable baseline
ripples which made a proper data reduction for weak lines impossible.
Fortunately, DAB turned out to be of little commercial interest, and it was
switched off. Further interference was found to be caused by local equipment
(workstations and active components). Most of the interference from such
components was removed either by shielding or switching off the devices.
\par Data taken before Sep 1996 generally are of good quality, with baselines
that were relatively easy to determine and not affected by interference. The Sep
1996 data vary in quality, as during some times of the day interference popped
up. Many spectra were unusable because of interference spikes at frequencies of
interest or because interference caused bad baselines. About 50\% of the
June/July 1999 data are generally usable, though some caution is needed.
Baselines may show strong curvature, and S-shaped interference spikes can be
seen in many spectra (see e.g.\ those of BD+10\,2179, HD\,83206, Mrk\,279,
PG\,0804+761 and PKS\,0837$-$12). In contrast with earlier runs, the spectra
from the two receivers (one for each polarization) were separately reduced, as
in about 20\% of the cases only one of the two was corrupted. Spectra taken in
2000 tend to show less interference and are of slightly higher quality than the
1999 set. However, again only about 50\% are usable.
\par In the 1995 and 1996 runs, a total of 469 spectra were taken, toward 173
different probes. Directions were selected in three ways. For about 40 probes
high-resolution absorption line data already existed. For about 130 probes
high-resolution absorption line data were anticipated or hoped for (about 50 of
these have now indeed been observed). The corresponding \HI\ data will be
published together with the absorption-line data. Finally, for 12 probes
high-angular resolution (1\to2 arcmin) Westerbork data existed, and a 3$\times$3
grid of positions was observed.
\par During the 1999 and 2000 runs 247 spectra were taken of 105 different
probes. All of these were selected because high-resolution absorption line data
existed (see Paper~I). For this set the results for each polarization were
separately reduced, effectively leading to 494 spectra. Further, integration
times for individual spectra were limited to 5 minutes, even if the total on the
target could be up to 30 minutes. Each of these spectra was judged individually,
and if interference clearly posed a problem the spectrum was eliminated. The
remaining spectra were averaged (a few times also adding in a spectrum with
shorter integration time from 1995 or 1996) to form the final spectrum toward a
target.
\par Because of some overlap between the two sets of runs (for targets where the
earlier spectra were not good or deep enough), a total of 249 probes has been
observed. Of these, 91 have high-resolution absorption data. The \HI\ column
densities for these are included in the table in Paper~I \dash\ the spectra are
shown here.

\subsection{Data reduction and calibration}
The data processing has been performed using the standard reduction programs of
the Max-Planck-Institut f\"ur Radioastronomie and the Radioastronomical
Institute of the University Bonn. The antenna temperature of the observed
spectra has been calibrated against IAU standard position S7 (Kalberla et al.\
1982). Afterwards the stray radiation from the sidelobes of the antenna diagram
was removed (Kalberla et al.\ 1980).
\par The instrumental baseline was removed after fitting a third order
polynomial to those channels which were free of line emission. A number of
observations were seriously affected by baseline ripples and a polynomial of
order 5 to 7 had to be used. This was the case for e.g.\ the spectra of
HD\,12323, HD\,12993 and PG\,1519+640, where a residual ripple can be seen. In
cases such as these, the brightness temperatures for the low-velocity gas are
unreliable, but the baseline under the HVC is still acceptable for determining
N(\HI) to within $\sim$25\%. If this is the case and if the star is not critical
for determining a lower distance limit to a cloud, the spectrum is still
included in the sample.
\par The spectra were taken with two HEMT receivers for two independent branches
with different circular polarization. The data processing was independent for
both channels. After the reduction we intercompared and averaged both spectra.
Interference and instrumental problems usually resulted in discrepant spectra.
Such data have been rejected except for those cases in which obviously only a
single channel was affected.
\par The final rms noise varies with the integration time and also somewhat
between runs. A typical value is 0.055\,K in 10 minutes for the 1995/1996 data.
For the 1999/2000 data interference and baseline problems cause the measured
noise to vary by a factor up to 1.5 for a given integration time, although the
average value still is typically 0.055\,K in 10 minutes.

\section{Other Observations}
\subsection{Overview}
For slightly over half of the probes in Paper~I, N(\HI) is based on data from a
telescope other than Effelsberg. This can be for one of four reasons. a)
Higher-angular resolution data are available (23 probes; sometimes these were
combined with the Effelsberg spectrum). b) Higher-quality (i.e.\ higher
signal-to-noise ratio or less interference) single-dish spectra are available
(44 probes). c) The probe is in the southern sky (14 probes). d) No directed
single-dish data are available (67 probes). In the latter case the data in the
Leiden-Dwingeloo Survey (LDS, Hartmann \& Burton 1997) were used. This gives
spectra at each half-degree point in longitude and latitude. A weighted average
was constructed from the four spectra surrounding the target direction, with
weights max(0,0.5\to$\sqrt{\Delta l^2 + \Delta b^2}$) (with $\Delta l$ and
$\Delta b$ in degrees).
\par Special cases are posed by 31 targets for which Effelsberg data were used
to derive N(\HI), even though Fig.~1 does not show the Effelsberg spectrum in
the direction of the probe. First, for the 22 stars in M\,15 for which
absorption-line data exist, the final value of N(\HI) is based on an
interpolation between nine Effelsberg spectra that are positioned on a
3$\times$3 grid with 5\arcmin\ spacing. Only the central position is shown in
Fig.~1. Second, for 9 probes the published value is based on an Effelsberg
spectrum, but the original data were lost. This pertains to 3 probes of
complex~A observed by Lilienthal et al.\ (1990), and to 6 probes of complexes C,
H and K observed by Centuri\'on et al.\ (1994). For these targets, Fig.~1 shows
a spectrum based on the Leiden-Dwingeloo Survey (see below), but the best final
value of N(\HI) is based on the published spectrum.
\par There are also a number of cases where a probe was observed at Arecibo or
with an interferometer as well as at Effelsberg, but we do not have the
high-angular resolution spectrum available in digital form. Then the Effelsberg
spectrum is shown in Fig.~1, but the label contains, e.g., ``Use Arecibo'' as a
warning that the column density used in Paper~I was derived from the better
data.
\par A final special case is presented by SN\,1993\,J. Two spectra are shown in
Fig.~1. First, the actual spectrum from de Boer et al.\ (1993), which includes
emission from M\,81 between velocities of $-$250 and +150\,\kms. Second, a
spectrum in which M\,81 was removed by means of a third-order polynomial fit.

\subsection{Spectra with angular resolution $<$9\arcmin}
For 9 probes the Effelsberg spectrum is complemented by data taken with the
Westerbork Synthesis Radio Telescope (WSRT): 0159+625 from Wakker \& Schwarz
(1991), PG\,0832+675 from Schwarz et al.\ (1995), Mrk\,106, PG\,0859+593,
PG\,0906+597 from Wakker et al.\ (1996), HD\,135485 and 4\,Lac from
Stoppelenburg et al.\ (1998), Mrk\,290 from Wakker et al.\ (1999) and Mrk\,205
from Braun \& Burton (2000). In this case Fig.~1 shows the Effelsberg spectrum,
but Paper~I uses the value of N(\HI) derived from the combination of WSRT and
Effelsberg data, as indicated in the label (see e.g.\ the plot for 0159+625).
\par For two probes (NGC\,3783 and HD\,101274) Paper~I gives N(\HI) from a
combination of data from the Australia Telescope Compact Array ({\it ATCA}) and
the Parkes telescope (Wakker et al.\ 2001). Fig.~1 shows the Parkes spectrum.
\par For the 10 stars that have been observed in M\,13 (Shaw et al.\ 1996),
Paper~I gives N(\HI) based on a combination of data from Jodrell Bank
(12\arcmin\ beam) and the Dominion Radio Astronomy Observatory ({\it DRAO},
1\arcmin\ beam). Fig.~1 shows a spectrum based on the Leiden-Dwingeloo Survey.
\par Payne et al.\ (1978, 1980) and Colgan et al.\ (1990) used Arecibo
(3\arcmin\ beam) to derive values for N(\HI) in the direction of seventeen 21-cm
radio continuum sources, in whose spectrum they searched for \HI\ absorption.
Arecibo was also used by Tamanaha (1996) to derive N(\HI) toward the probes in
core AC0. For these targets, Fig.~1 shows the LDS spectrum, but the Arecibo
value for N(\HI) is used in Paper~I, as indicated in the label (see e.g.\
3C\,78).

\subsection{Higher-quality single-dish spectra}
High-quality spectra for a number of probes of complex~A, M and the IV-arch were
obtained by Ryans et al.\ (1997a, 1997b) using the Jodrell Bank telescope
(12\arcmin\ beam). For most of these an Effelsberg spectrum was also obtained,
except for PG\,1213+456 and H.O.+41B. Further, for BD+38\,2182, HD\,93521,
HD\,203664 and HD\,205556 the Effelsberg data are of much lesser quality than
the Jodrell Bank spectrum. For the latter two, Fig.~1 shows the Effelsberg
spectrum, but for the other four probes, Fig.~1 shows the Jodrell Bank spectrum,
courtesy of R.\ Ryans. For all six targets the value of N(\HI) derived from the
Jodrell Bank spectrum is used in Paper~I.
\par In 1992, at our request, C.\ Heiles used the Hat Creek telescope
(36\arcmin\ beam) to make an especially deep observation (24 hours integration
time) of the direction toward SN\,1991\,T. In this direction Meyer \& Roth
(1991) had detected weak \CaII\ absorption at velocities of +215 and +263\,\kms.
Even though the Hat Creek \HI\ spectrum has a detection limit of
3\tdex{17}\,\cmm2, no emission is seen at these velocities. Fig.~1 shows the Hat
Creek spectrum.
%VALUES: anomabs L x; egrep 'DLMS|ABML' ABSLINES.tex | awk -F\& '{print $1,$9,$13,$23}' | sort | uniq
\par Green Bank 140-ft data (21\arcmin\ beam) were used by Danly et al.\ (1992)
and Albert et al.\ (1993) in their absorption-line studies of IVCs. Paper~I
includes 16 of their stars, for 10 of which an improved spectrum was obtained
using Westerbork, Effelsberg or Jodrell Bank; for 4 stars the LDS spectrum is
used instead of the Green Bank data. HD\,86248 lies at too low a declination for
both northern telescopes, but the Parkes data are not yet properly reduced to
improve on the Green Bank spectrum. Further, for the +73\,\kms\ component toward
HD\,100340 and the $-$43\,\kms\ component toward HD\,137569, the published
column density limit based on the Green Bank spectrum is better than the limit
based on the LDS spectrum.
\par Savage et al.\ (2000) used previously unpublished data obtained by Murphy
with the Green Bank 140-ft telescope in the direction of many AGNs that were
also observed using the Faint Object Spectrograph on the Hubble Space Telescope.
These spectra are of very high quality and have noise levels of about 0.015\,K.
For a few probes a high-quality Effelsberg spectrum also exists (3C\,351.0,
H\,1821+643, HS\,0624+6907, PKS\,1136$-$13 and PG\,1116+215). The first of these
samples the edge of an HVC, which is detected in the larger Green Bank beam, but
not in the smaller Effelsberg beam. For the other four Savage et al.\ (2000) do
not list the HVC component separately, but only mention that the \HI\ spectrum
has a high-velocity tail. Thus, for these five probes the value of (or limit
for) N(\HI) is derived from the Effelsberg spectrum, while for the remaining
AGNs discussed by Savage et al.\ (2000) Paper~I uses the values for N(\HI) and
velocity derived from the Green Bank spectrum. In Fig.~1 the Effelsberg spectrum
is shown for 3 of these, while the LDS spectrum is shown for the 10 others.

\subsection{Southern targets}
Sixteen HVC/IVC probes lie below declination $-$35\deg, the limit of the
Leiden-Dwingeloo Survey. For two of these an {\it ATCA} spectrum was obtained
(NGC\,3783 and HD\,101274, see above). For the remaining probes, spectra were
extracted from observations made with the 64-m Parkes radio telescope equipped
with a 21-cm multi-beam receiver (Staveley-Smith 1997). For ten of these,
high-velocity resolution data (0.8\,\kms) were available via observations made
with the narrow-band multi-feed facility (Haynes et al.\ 1999; Br\"uns et al.\
2001). The narrow-band system utilizes the central 7 beams of the 13-beam
multibeam receiver with two orthogonal polarizations and a bandwidth of 8 MHz.
Narrow-band data were unavailable for three of the sources, and the spectra were
therefore taken from the HVC-reduced (Putman et al.\ 2001) ``\HI\ Parkes All-Sky
Survey'' ({\it HIPASS}; Barnes et al.\ 2001). The standard {\it HIPASS}
reduction method filters out emission which extends over more than 2\deg\ in
declination, while the HVC method recovers the emission unless it completely
fills an 8\deg\ scan. After Hanning smoothing, these data have a velocity
resolution of 26\,\kms\ and this low resolution makes it difficult to discern
the weak HVC (and especially IVC) components. At 21-cm the Parkes telescope has
a beamwidth of 14\arcmin, but the gridding process which combines the individual
scans increases the spatial resolution to 15\farcm5$\pm$1\arcmin. The spectra
were extracted from the gridded data at the pixel closest to the coordinates of
the probe.

\section{Component fitting}
To derive column densities for the HVC and IVC components in the spectra, a
decomposition into gaussians was made. The fitting procedure requires an
estimate of the rms noise and a set of initial estimates. Since the purpose of
the fitting is to determine N(\HI) and \vlsr\ for the HVCs and IVCs, no great
effort was made to obtain perfect fits to the peak near 0\,\kms\ that is present
in all spectra. Still, that component is often fitted quite well, although in a
few cases line wings were not fitted. For the HVC and IVC components, however,
extra care was always taken. For probes near the Galactic plane (in HVC
complex~H), the low-velocity gas was not fitted, as the structure is very
complex and components can easily be artificial superpositions in the line of
sight.
\par The rms was estimated from a spectral region free from \HI\ emission.
Wherever possible the region between $-$400 and $-$200\,\kms\ ($-$250\,\kms\
when necessary) was used. However, for a substantial fraction (33\%) of spectra
there is either an HVC component (10\% of the 33\% lie in the Magellanic Stream)
or interference at those velocities, in which case an alternative range is used.
See Fig.~1 for the actual selected velocity range for each target.
%RMS   -400 -200  132
%RMSX  -400 -250   31
%RMSx    50  200    7
%RMSp   100  300   19 (10 MS)
%RJB1  -190 -125    4
%RJB2  -250 -150    0
%RMSP  -250 -100    7 (7 MS)
%other             40 (5 MS)
%:MS:              25 (10 RMSp 7 RMSP)
\par For about half of the spectra it was possible to make an initial estimate
automatically. For the other half, it was necessary to help the fit along by
giving a fairly good estimate of the amplitude, velocity and width. Care was
taken to make sure that in general the final fitted components were no broader
than about 40\,\kms. However, there are a few exceptional cases where the HVC or
IVC profile is fitted just as well by a single component as by two components
(e.g.\ for I\,Zw\,18). Larger widths were then allowed.
\par In 40\% of the cases the fit was made piecewise, i.e.\ two spectral regions
were fit separately. This mostly happens when the HVC component is weak. Then
the formal fit for the low-velocity component can usually be improved by adding
one (or more) components, even if these are not clearly seen by eye. The fitting
procedure then tends to ignore the weak HVC component, unless an unrealistic
total number of components is used. Sometimes piecewise fits were also needed if
the central component has broad wings. Free fitting then tends to create
components with unrealistic FWHMs of 80\,\kms\ or more, which then absorb the
HVC/IVC component. By restricting the velocity range of the fit this problem can
usually be avoided. This procedure works well in most cases, but a few times
some artifacts remain visible in the fit for the low-velocity gas (e.g.\ for
3C\,395, HD\,32641, HD\,45315 and Mrk\,509).

\section{Results}
Figure~1 shows the spectra and the fits. All probes in Table~2 of Paper~I are
shown. The order is alphabetical, rather than by cloud, as about a third of the
sightlines goes through more than one cloud.
\par For each target the \HI\ spectrum is shown with two vertical scales. The
top scale emphasizes the low-velocity gas, the bottom scale gives the clearest
view of the HVC or IVC component. In both cases the gaussian fit is superposed
on the actual spectrum. The velocity range used to estimate the rms noise is
shown as a horizontal bar in the top spectrum. A label is included too, which
gives on the first line the probe name and on the second line the galactic
longitude and latitude. The third line gives some information about the
observations. For spectra obtained using Effelsberg, the date of observation is
given in the format ``yymmdd'', as is the integration time, in minutes. If the
final N(\HI) values used in Paper~I were obtained from another telescope this is
indicated too, by a label such as ``Use Arecibo''. The fourth label line shows
the velocity resolution and rms noise. The velocity resolution is 1.28\,\kms\
for Effelsberg data, 1.21\,\kms\ for Jodrell Bank data, 0.82\,\kms\ for Parkes
data and 1.03\,\kms\ for spectra extracted from the LDS. For all these the label
gives ``1 km/s''. For Parkes {\it HIPASS} data the velocity resolution is
26\,\kms. In a number of cases it was necessary to smooth to 2, 4 or even
8\,\kms\ to be able to fit a weak HVC component. The lower resolution data are
then shown on the plot.
\par Below the two spectra a list of components is given. This gives for each
component the central velocity (in \kms), the amplitude (in K), the FWHM (in
\kms) and the derived column density with its error (in units of
\dex{18}\,\cmm2). The final column gives (for HVC and IVC components) the cloud
name with which the component is identified, based on the catalogues of Wakker
\& van Woerden (1991) and Kuntz \& Danly (1996) and some new definitions given
in Paper~I. In a few cases the text ``Other IVC'' is given, which refers to an
unnamed IVC listed in the set of ``Other negative/positive IVCs'' in Table~2 of
Paper~I. Notes in square brackets, such as ``[WSRT: $-$199 190]'' give the
telescope, velocity and column density for HVC and IVC components if the final
value was based on better data than the Effelsberg or LDS spectrum that is
shown. In a few cases, such a note is not associated with a fitted component,
indicating the existence of an absorption without associated \HI, or the
existence of an \HI\ component fainter than what is visible in the displayed
spectrum.
\par For a number of spectra the listed component has a large systematic error
(i.e.\ much larger than the listed statistical error). There can be various
reasons for this. The most important of these cases are individually described
below.
\par 1) 3C\,351.0, complex~C component at $-$130\,\kms. The fits give
N(\HI)=4.5$\pm$0.6\tdex{18}\,\cmm2\ for the 35\arcmin\ LDS beam (see Fig.~2a for
the spectrum), 2.8$\pm$0.1\tdex{18}\,\cmm2\ for the 21\arcmin\ Green Bank beam
(see Lockman \& Savage 1995 for the spectrum) and 4.2$\pm$0.3\tdex{18}\,\cmm2\
for the 9\arcmin\ Effelsberg beam (see Fig.~2a and Fig.~1b). In the LDS spectrum
this component looks particularly noisy and the fitted width may be too large.
Similarly, in the Green Bank spectrum, the component is particularly faint, and
is not a nice easy-to-fit gaussian. Finally, the Effelsberg spectrum may have
been affected by interference. Thus, the systematic errors are large. We
estimate them to be $\sim$1.5\tdex{18}\,\cmm2. Nevertheless, it also may be the
case that there is much structure in the cloud. In the neighbouring LDS profiles
the $-$130\,\kms\ component varies by a factor of 2. Thus, The Effelsberg beam
might be picking up a brighter spot, which is beam diluted in the Green Bank
beam, whereas multiple bright spots are seen in the LDS beam. Better data is
needed to settle this question.
\par 2) For 3C\,418 six components are needed to fit the profile between $-$150
and $-$50\,\kms. However, they are badly blended and probably incorrect in
detail. The combined N(\HI) for the three Outer Arm components is probably OK to
within 20\%, however.
\par 3) The complex~G component at $-$107\,\kms\ fitted for the 4\,Lac sightline
is rather wide, not symmetrical and probably a blend.
\par 4) The parameters of the two IVCs at $-$79 and $-$42\,\kms\ fitted to
BT\,Dra are rather uncertain, as they blend with the HVC and the low-velocity
gas. The systematic error is estimated to be $\sim$2\tdex{18}\,\cmm2, or 50\%.
\par 5) Toward H\,1821+643, there are substantial differences between the LDS,
Effelsberg and Green Bank spectra. In the LDS spectrum (Fig.~2b) emission at
velocities $<$$-$80\,\kms\ is barely discernable (total column density estimated
at $\sim$1.2\tdex{18}\,\cmm2). In the Green Bank spectrum (Lockman \& Savage
1995) there is a broad component running from $-$140 to $-$80\,\kms, with total
column density $\sim$14.5\tdex{18}\,\cmm2, while the Effelsberg spectrum clearly
shows two components at $-$128 and $-$87\,\kms\ with a total column density of
11.8\tdex{18}\,\cmm2. A larger-scale look at the LDS data reveals that
H\,1821+643 sits in a hole in the large area of emission associated with the
Outer Arm. Apparently there is a small patch of bright \HI\ right around
H\,1821+643, which is picked up by the Green Bank and Effelsberg telescopes, but
beam-diluted in the LDS beam. However, the maximum dilution factor between the
21\arcmin\ Green Bank and the 35\arcmin\ LDS beam is only a factor 3, so some of
the discrepancy must be due to a baseline error or just too low an S/N ratio in
the LDS spectrum.
\par 6) The $-$70\,\kms\ IV spur component toward HD\,103400 blends with the
lower-velocity gas. The real uncertainty in the column density is probably twice
the given statistical error.
\par 7) The $-$100\,\kms\ complex~L component toward HD\,135485 is four times
weaker in the 1\arcmin\ beam combined WSRT and Jodrell Bank data than in the LDS
data. This is probably related to the fact that the surrounding cloud is fairly
small and shows much structure.
\par 8) The +80\,\kms\ component toward HD\,203664 is very weak, and blends in
with the gas at less positive velocities. In the Jodrell Bank spectrum (Ryans et
al.\ 1996; N(\HI)=2.2$\pm$0.4\tdex{18}\,\cmm2) it is more clearly seen than in
the Effelsberg spectrum (Fig.~1; N(\HI)=1.1$\pm$1.5\tdex{18}\,\cmm2). Thus, this
component is clearly present, but its column density could be anywhere in the
range 0.5 to 2.5\tdex{18}\,\cmm2. The fits toward the nearby stars HD\,203699
and HD\,205556 (probing the same IVC) also suffer from problems with blending.
\par 9) The spectrum toward Mrk\,279 was fitted by 8 components, all of which
can be identified with continuous structures in the surrounding area. In this
case the fitted values appear fairly reliable, in spite of the blending.
\par 10) PKS\,2345$-$67, Magellanic Stream component at $-$174\,\kms. The listed
component is a fit to the left wing of the profile. At neighbouring positions
two separate components are seen at $-$190 and $-$130\,\kms, so the profile may
show a particularly bad blend of interpolated components.
\par 11) The parameters of the IVC fitted to the \HI\ spectrum toward
SN\,1981\,D are rather uncertain since the spectral resolution is just 26\,\kms.
\par 12) The VHVC fit to the \HI\ spectrum toward SN\,1986\,G is vary narrow.
This may or may not be realistic. Better data are needed to tell.
\par 13) For SN\,1993\,J, the contribution of M\,81 was removed by fitting a
third order polynomial to the velocity ranges $-$120 to $-$80 and +30 to
+70\,\kms. The LLIV and low-velocity component were then fitted. However, it is
not clear how much of the emission near $-$25\,\kms\ is associated with M\,81.
Including the velocity range $-$35 to $-$20\,\kms\ in the baseline reduces the
column density of the LLIV component from 80 to 60\tdex{18}\,\cmm2. The final
value used in Paper~I thus is 70$\pm$10\tdex{18}\,\cmm2.

\section{Analysis}
\subsection{Comparison of Leiden-Dwingeloo Survey and Effelsberg column densities}
For each of the directions in which an Effelsberg spectrum was available, we
also extracted a spectrum from the LDS in the manner described in
Sect.~\Sother.1. A gaussian fit was made to this spectrum, in almost all cases
using the same initial estimates as for the Effelsberg spectrum. In a few cases
subtle differences required an extra low-velocity component for one of the two,
or other small adaptations.
\par Figure~2 shows some examples of the differences observed in the profiles
with a 9.1 and a 35 arcmin beam. These examples were chosen to illustrate some
of the more extreme differences. The $-$181\,\kms\ component toward 3C\,351.0
illustrates that some faint HVC components can be clearly visible in a 35 arcmin
beam, but disappear at higher angular resolution. This spectrum also illustrates
structure in the low-velocity gas, which has N(\HI)=157\tdex{18}\,\cmm2\ at 35
arcmin, but 177\tdex{18}\,\cmm2\ at 9 arcmin. The spectrum of BD+63 985
illustrates fine structure in the IV arch (core IV21). Toward BS16079-0017 the
column density of the low-velocity gas increases by 30\% when going from a 35
arcmin to a 9 arcmin beam. The spectrum toward H\,1821+643 shows that
high-velocity components can also appear, rather than disappear, at higher
angular resolution. HD\,83206 illustrates that components can appear separated
at low angular resolution, but blended at higher angular resolution. The
spectrum toward M\,15 shows that small-scale structure also occurs at positive
velocities. Mrk\,106 illustrates another case of varying relative strength of
low- and intermediate-velocity gas. Toward Mrk\,116 the low-velocity component
at +7\,\kms\ is 60\% weaker at lower angular resolution. The 9 arcmin Effelsberg
spectrum of Mrk\,279 shows 8 identified components, all of which are present in
the LDS spectrum, but with different relative strengths. PG\,1259+593
illustrates how the high-velocity component differs by a factor 2 in column
density; this is near a core of HVC complex~C.
\par Using all of the gaussian fits, a table was made of N(\HI) observed at
either telescope, separately for the high-, intermediate- and low-velocity
components. The low-velocity peak usually had to be fit with 2 or 3 components,
once with 4, once with 5, while in 14\% of the cases 1 component was sufficient.
For the comparison, all low-velocity components were combined, as the exact
decomposition depends on the details of the profile rather than on the real
physical origin of the components.
\par Figures~3 and 4 present the results in graphical form. In Fig.~3 the left
panels show the ratio as function of log N(\HI-Eff), separately for high-,
intermediate- and low-velocity gas. The right panels show the histogram of
ratios for the three kinds of gas. Figure~4 shows the scatter plot and histogram
of column density differences.

\subsection{Results for HVCs and IVCs}
For high- and intermediate-velocity clouds the ratio, $R$, of N(\HI) measured
with a 9 vs a 35 arcmin beam lies in the range 0.25\to2.5 (0.3\to2.1 when
excluding the three most extreme points at either end). The full range is seen
in most of the individual HVCs and IVCs, suggesting that the small-scale
structure has similar characteristics in each of these. In one case
(PG\,0229+064, next to the Cohen Stream) the ratio is 0, as the HVC is not seen
at all in the Effelsberg beam, although the LDS spectrum suggests
N(\HI)=20\tdex{18}\,\cmm2. For HVCs the distribution of ratios peaks at a ratio
$R$$\sim$0.7 (the median is 0.8), and it is slightly skewed towards values of
$R$$<$1 (10 ratios fall between 0 and 0.5, 32 between 0.5 and 1, 18 between 1
and 1.5, and 9 above 1.5). For IVCs the distribution appears more symmetrical
(the median is 1.03 and 15 lie between 0.5 and 1, 18 between 1 and 1.5). The
skewness for the HVC ratio distribution may be real as bright condensations fill
only a fraction $<$50\% of the Dwingeloo beam, and an arbitrary direction will
in most cases have smaller N(\HI) than the average of a larger area seen in the
Dwingeloo beam.
\par For the HVCs and IVCs, the maximum difference between the two column
density measurements tends to increase with increasing N(\HI), and can go either
way. The histograms of differences are centered around no difference, with a
range of $\pm$30\tdex{18}\,\cmm2. This width is mostly determined by the
directions with high N(\HI), as the range of ratios is independent of N(\HI).

\subsection{Results for low-velocity gas}
For low-velocity gas, the ratio distribution is much narrower, with a sharp
peak and a range of only 0.76\to2.0 (0.92\to1.4 when excluding the three most
extreme points at either end). However, the peak is centered around a ratio of
1.1, not 1. Below we suggest that this offset is an artifact of the
stray-radiation correction. If so, the real range of ratios would be 0.85\to1.3.
This range is smaller than for HVCs and IVCs. The histogram of differences for
low-velocity gas is offset from zero by about 15\tdex{18}\,\cmm2, and positive
differences (i.e.\ N(\HI-Eff)$>$N(\HI-LDS)) dominate. This is the same effect as
seen in the offset of the average ratio.
\par There are several possible explanations for the difference between
HVCs/IVCs and low-velocity gas in the range of ratios. 1) There is intrinsically
less small-scale structure in the nearby low-velocity gas. 2) The relative
magnitude of fluctuations may decrease with spatial size, and because the
low-velocity gas is closer we are probing smaller absolute scales. 3) This is an
artifact because for the low-velocity gas two or more components are mixed
together, so that upward fluctuations in one component may be compensated by
downward fluctuations in another component. That the third of these effects
plays a role is indicated by the histogram of ratios for the 11 sightlines where
just one component was fit for the low-velocity gas. That distribution has a
larger dispersion than the histogram for all low-velocity gas.
\par The offset in the average value of N(\HI-Eff) and N(\HI-LDS) observed for
the low-velocity gas is not completely understood. It is unlikely to be due to a
calibration (i.e.\ scaling) error, as then the same offset should have been seen
for the HVCs and IVCs. A more likely possibility is that the stray-radiation
correction is not completely correct. This correction has been applied to both
datasets. The LDS data are used to correct themselves, and the Effelsberg data
are corrected using the corrected LDS-data. Apparently there is a slight
undercorrection in the latter case, which amounts to about 15\tdex{18}\,\cmm2\
on average. For the high-latitude directions studied here that is on average
10\% of N(\HI). However, most of the correction is due to higher column density
low-latitude directions, so this is no more than $\sim$1\% of the total
correction.
\par Such a problem may indicate that the correction for the far sidelobes is
slightly incorrect (about 1 to 2\% of the total correction). The Effelsberg
antenna pattern was determined by Kalberla et al.\ (1980) and updated in 1992.
After this date some mechanical changes within the telescope surface were made.
A problem may arise from spoilers which have been mounted on the feed support
legs. The intention was to eliminate baseline ripples due to standing waves at
short wavelengths. However, at 21 cm these spoilers are expected to increase the
effective cross section for the scattering of waves falling on the feed support
legs. We suspect that the stray cones of the antenna diagram due to the feed
support legs have been altered by these mechanical modifications. Attempts to
improve the quality of the antenna diagram which is used to correct for stray
radiation remained unsuccessful.

\subsection{Comparison with column densities measured at high resolution}
Table~1 lists the values measured for N(\HI) in the direction of 16 probes of
HVCs for which \HI\ data exist at high resolution. There are two groups. First,
probes for which 21-cm interferometer data exist, so that individual velocity
components can be compared with single-dish data. Second, distant stars or
extra-galactic probes for which the integrated \HI\ column density along the
line of sight was also measured using Ly$\alpha$ absorption.
\par This table shows that the ratio N(\HI-2\arcmin)/N(\HI-35\arcmin) has the
same wide range of 0.25\to2.5 as the ratio N(\HI-9\arcmin)/N(\HI-35\arcmin).
When excluding the extreme value at each end the range is 0.54\to2.0. However,
the ratio N(\HI-2\arcmin)/N(\HI-9\arcmin) lies in the much narrower range
0.74\to1.24 (0.75\to1.06 when excluding the extreme value at each end). This is
consistent with the result described in Sect.~\SanalHVC, where we showed that
the contrast between a 9 and a 35 arcmin beam is a factor up to 2.5. The
difference in ranges suggests that N(\HI) derived from a 9\arcmin\ beam
approximates the value obtained at higher resolution to within 25\%, and that on
average N(\HI) tends to be slightly lower at higher resolution.
\par Measuring N(\HI) using Ly$\alpha$ absorption gives the smallest possible
beam ($<$0.1 arcsec). However, only the combined column density in all
components can be measured. Still, this may give some indication of structure on
scales smaller than the 1\arcmin\ limit of interferometers, since the integrated
column density is usually dominated by the low-velocity components. Effelsberg
data exist toward 6 AGNs in the sample of Savage et al.\ (2000), who measured
N(\HI-Ly$\alpha$) using the Faint Object Spectrograph ({\it FOS}) on the Hubble
Space Telescope. Effelsberg data are also available for distant two stars in the
sample of Diplas \& Savage (1994), who measured N(\HI-Ly$\alpha$) using {\it
IUE}. Excluding the deviant value toward HS\,0624+6907, which has a particularly
large systematic uncertainty, the ratio N(\HI-Ly$\alpha$)/N(\HI-Eff) lies in the
range 0.74\to0.98. This range is slightly narrower but similar to that found for
N(\HI-Ly$\alpha$)/N(\HI-Green Bank) (0.63\to0.87), but smaller than that for
N(\HI-Ly$\alpha$)/N(\HI-LDS) (0.69\to1.24).
\par Even though this is a small sample, it suggests that measurements of N(\HI)
with a 9 arcmin beam are sufficient to obtain reasonable column density
estimates, whereas measurements with a 35 arcmin beam are only accurate to a
factor of about 2\to3.
\par It would be useful to also compare the Effelsberg 9 arcmin and LDS 35
arcmin column densities with observations with an intermediate beam, such as the
21 arcmin beam provided by the Green Bank 140-ft. The data of Murphy et al.\
(2000) for AGNs observed with the {\it FOS} provide numbers for about 190
directions. However, so far only 10 of these have also been observed at
Effelsberg, and most do not show well-isolated HVC or IVC component. So, only a
limited comparison will be possible.

\section{Conclusions}
\par a) For HVCs and IVCs improving the resolution by a factor $\sim$4 (from
35\arcmin\ to 9\arcmin\ beam, a factor 15 in area) can change the derived value
of N(\HI) by a factor up to about 3 either way. In most (two-thirds) of the
directions the change is a factor $>$1.5 either way, while in just one-third is
the change a factor between 0.5 and 1.5.
\par b) For HVCs/IVCs where higher resolution data are available, the ratio
between N(\HI) as observed with a 1\arcmin\to2\arcmin\ beam vs a 9\arcmin\ beam
(a factor $\sim$20 in area) lies in the range 0.74\to1.24 whereas comparing
1\to2\arcmin\ data with 35\arcmin\ data (another factor $\sim$20 in area) gives
a range 0.25\to2.5.
\par c) For sightlines where N(\HI-total) was derived from Ly$\alpha$ absorption
the ratio N(\HI-Ly$\alpha$) to N(\HI-9\arcmin,total) lies in the range
0.74\to0.98.
\par d) Combining b) and c) suggests that a 9\arcmin\ beam is sufficient to
derive abundances to within about 25\%, whereas using a larger beam leads to
larger uncertainties, of up to a factor 3.
\par e) On average the single-dish column densities tend to be slightly higher
than the values measured at higher resolution, so that on average ionic
abundances tend to be underestimated by 0\to25\%.

\acknowledgements
The Effelsberg Telescope belongs to the Max Planck Institute for Radio Astronomy
in Bonn. We thank Carl Heiles for observing the sightline to SN\,1991\,T for 24
hours, using the Hat Creek telescope. Finally, we thank Robert Ryans for
providing his Jodrell Bank spectra.

%%%%%%%%%%%%%%%%%%%%%%%%%%%%%%%%%%%%%%%%%%%%%%%%%%%%%%%%%%%%%%%%%%%%%%%%%%%%%%%%

\def\fgnumber#1{\noindent Figure #1.\ } %AASPP

\def\InsertPage#1#2#3#4{\newpage\vbox to \vsize{\vss{\small#4}} %AASPP
\includegraphics{#1}} %AASPP

%%%%%%%%%%%%%%%%%%%%%%%%%%%%%%%%%%%%%%%%%%%%%%%%%%%%%%%%%%%%%%%%%%%%%%%%%%%%%%%%

\InsertPage{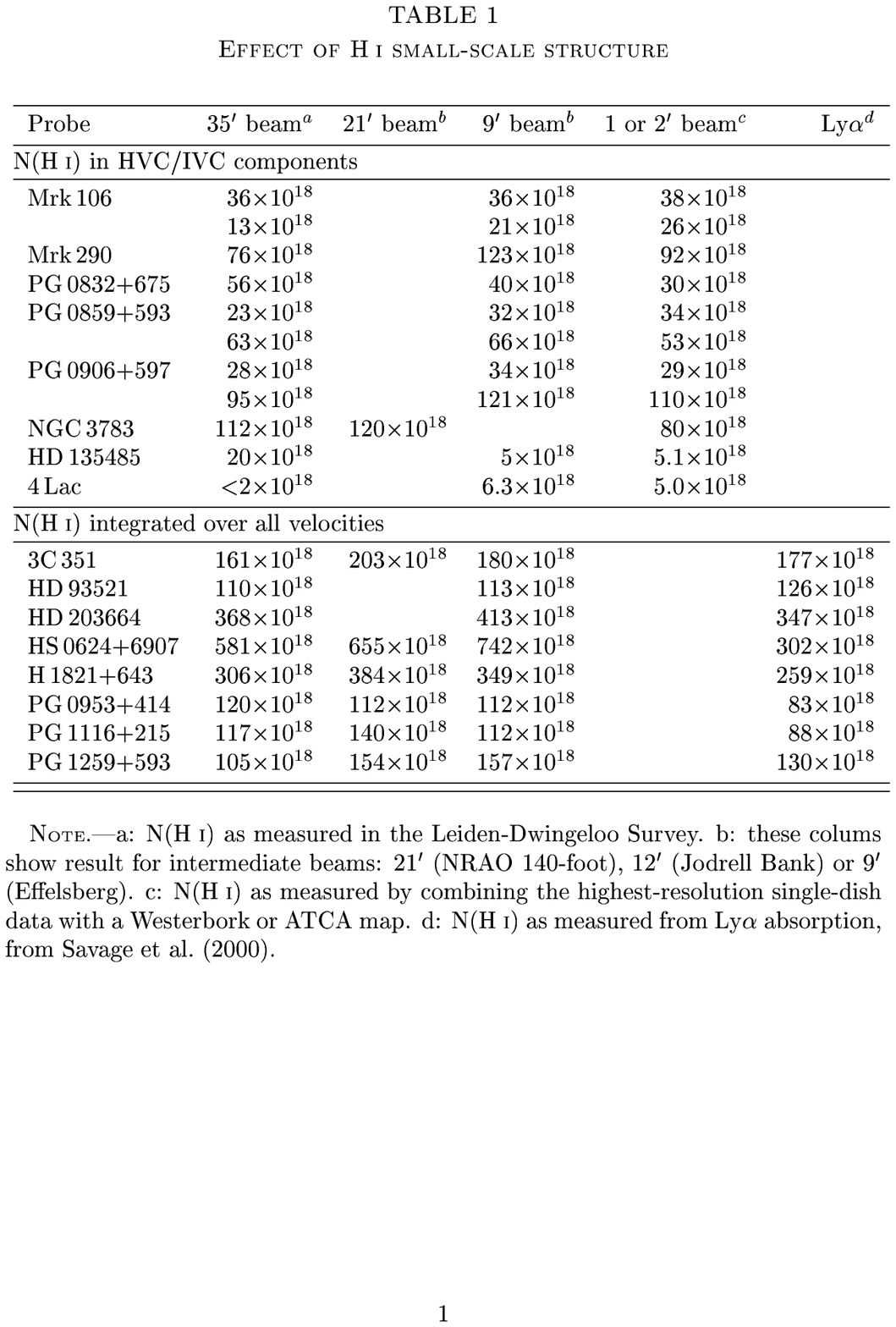}{-45}{-75}{\null}  %AASPP

\InsertPage{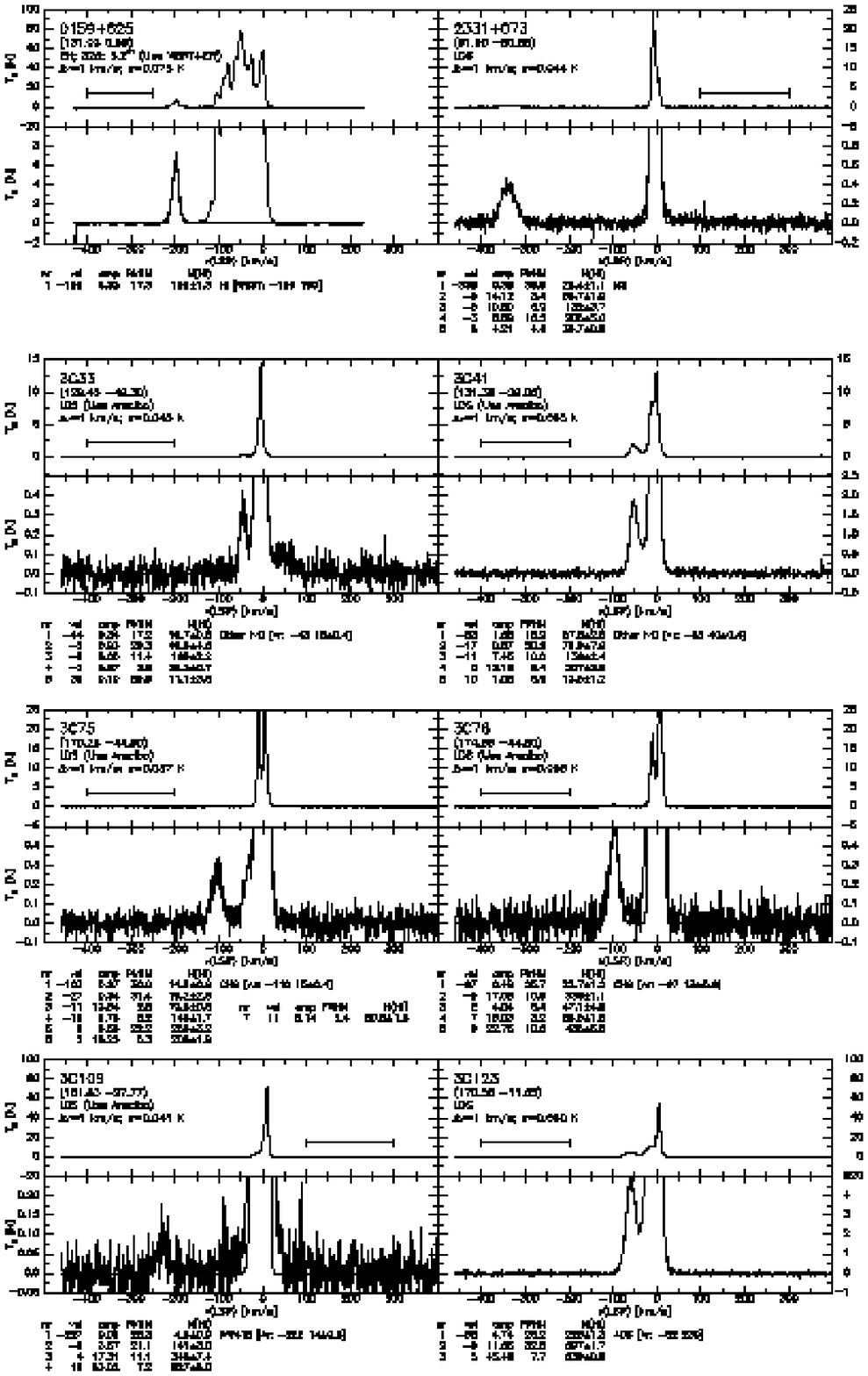}{-45}{-30}{% %AASPP
\fgnumber{1a} \HI\ spectra in sightlines toward absorption-line probes. A
detailed description is given in Sect.~\Sresults. Problematic components are
commented upon in that section.}
\InsertPage{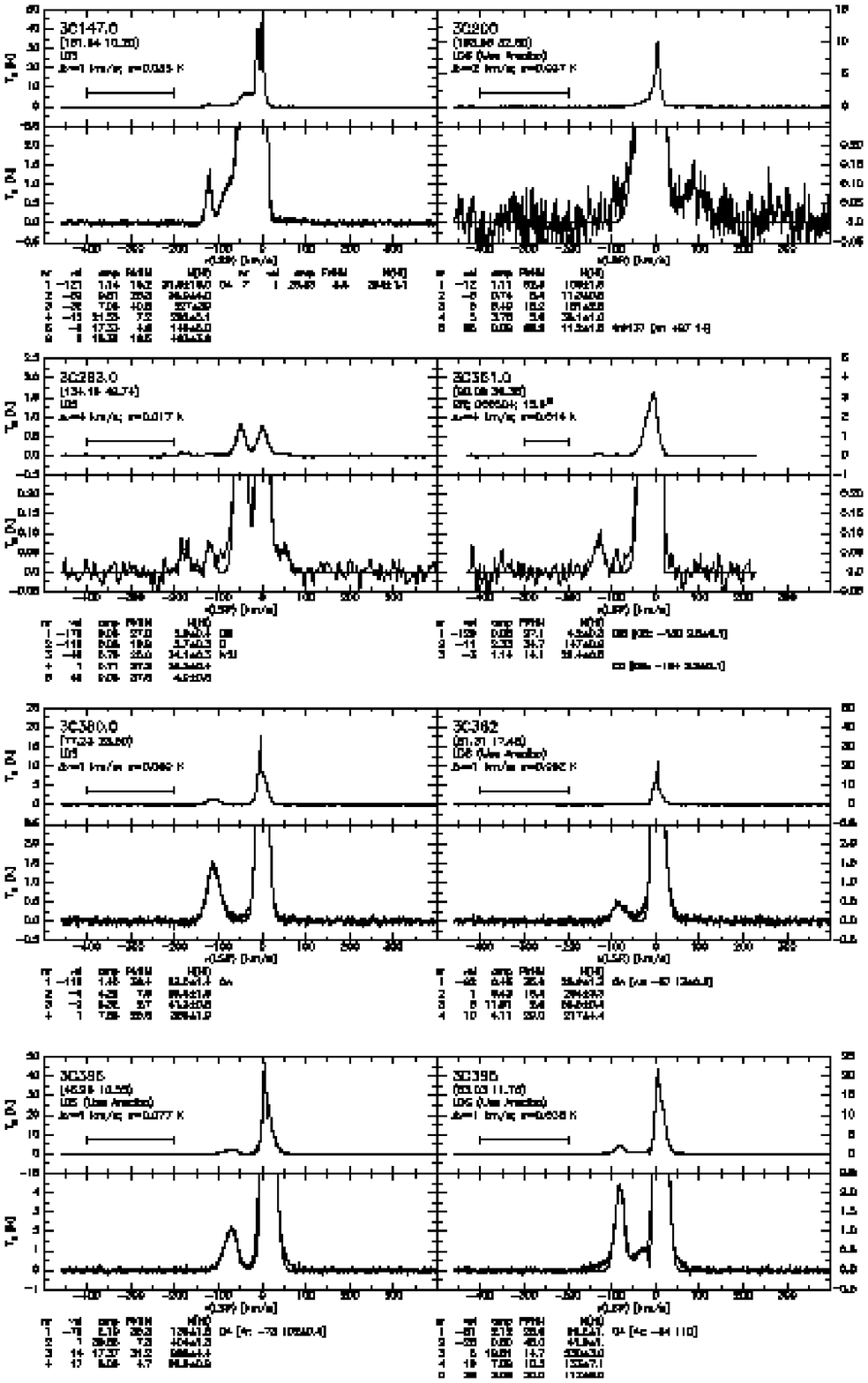}{-45}{-45}{Fig. 1b} %AASPP
\InsertPage{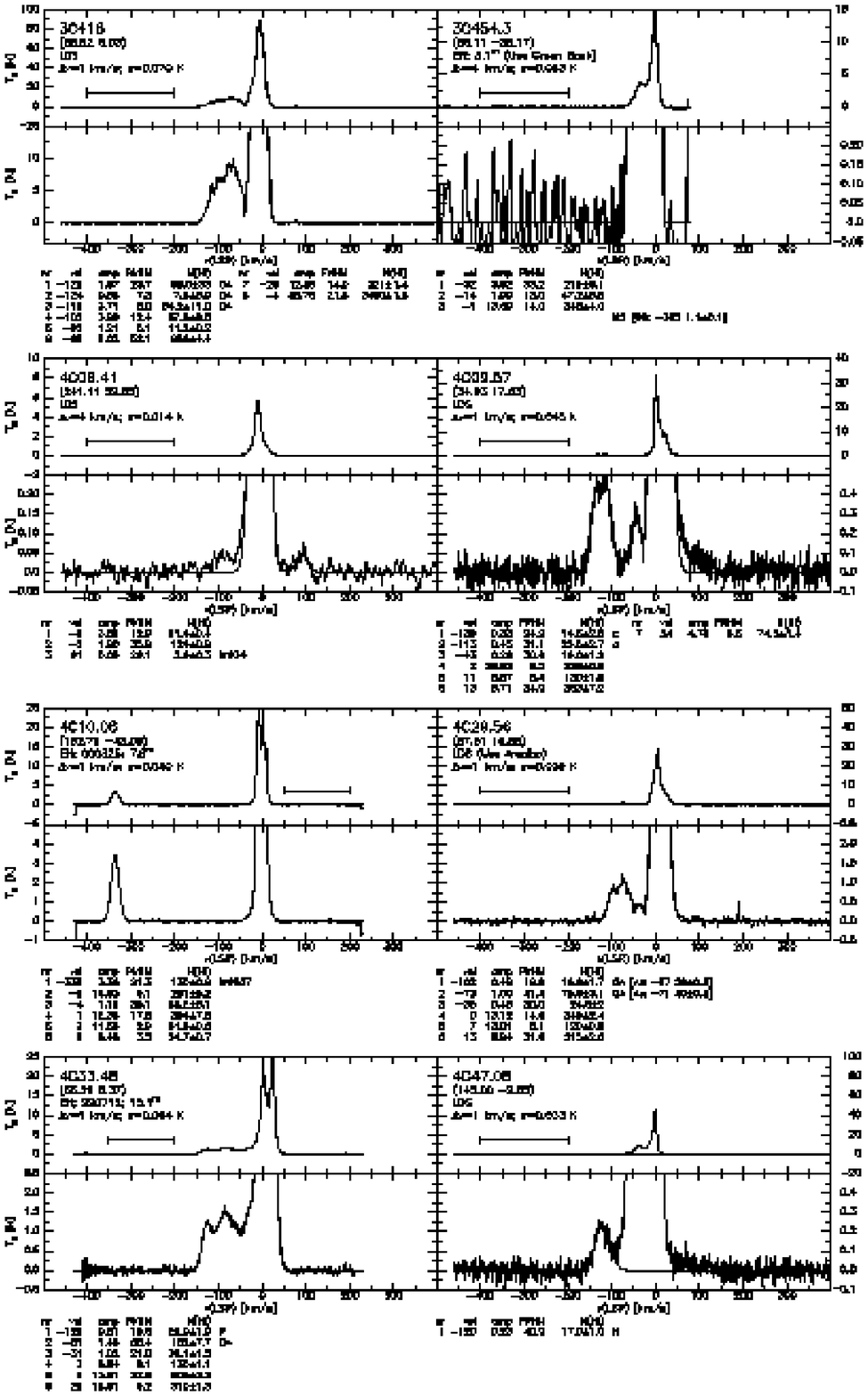}{-45}{-45}{Fig. 1c} %AASPP
\InsertPage{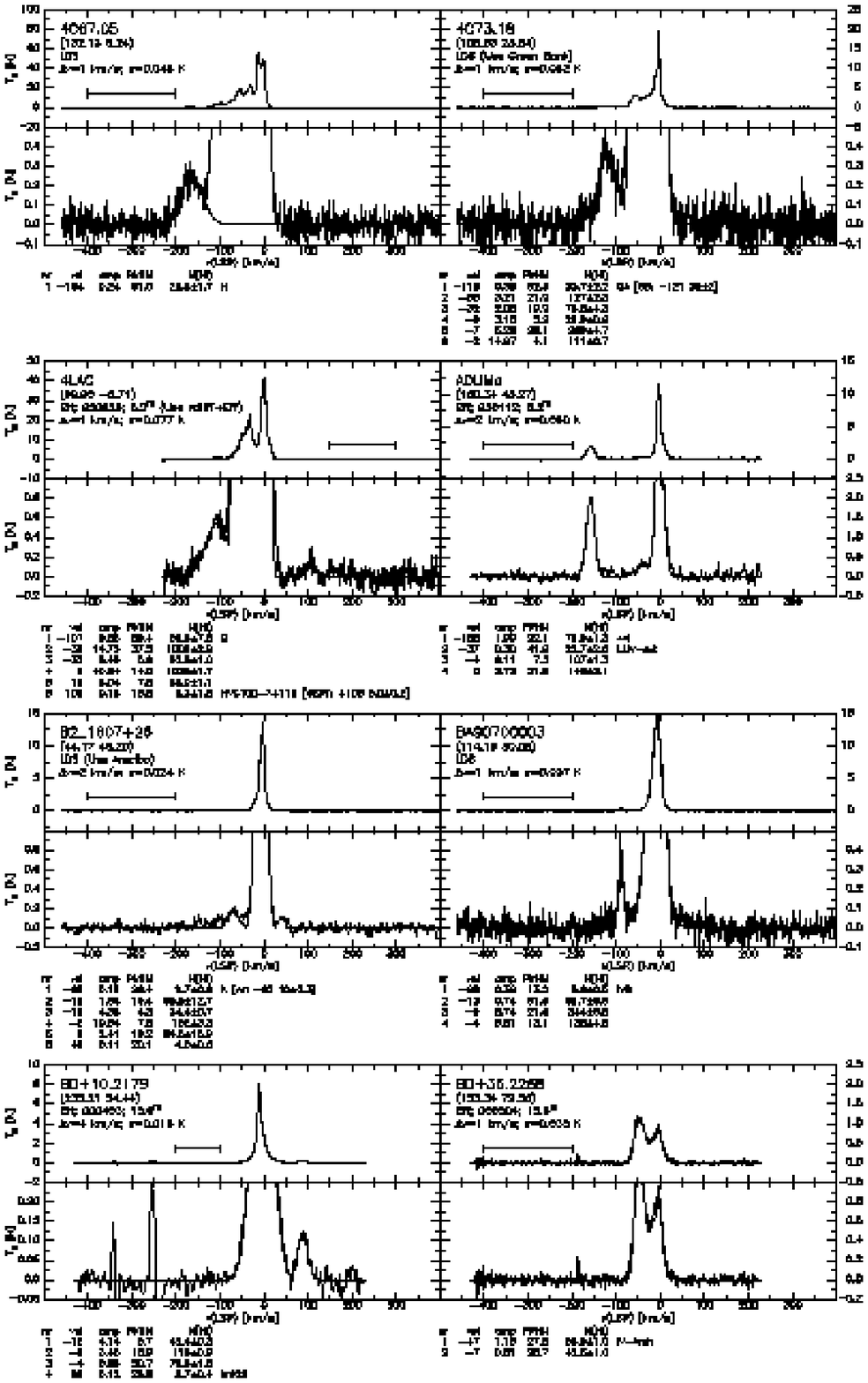}{-45}{-45}{Fig. 1d} %AASPP
\InsertPage{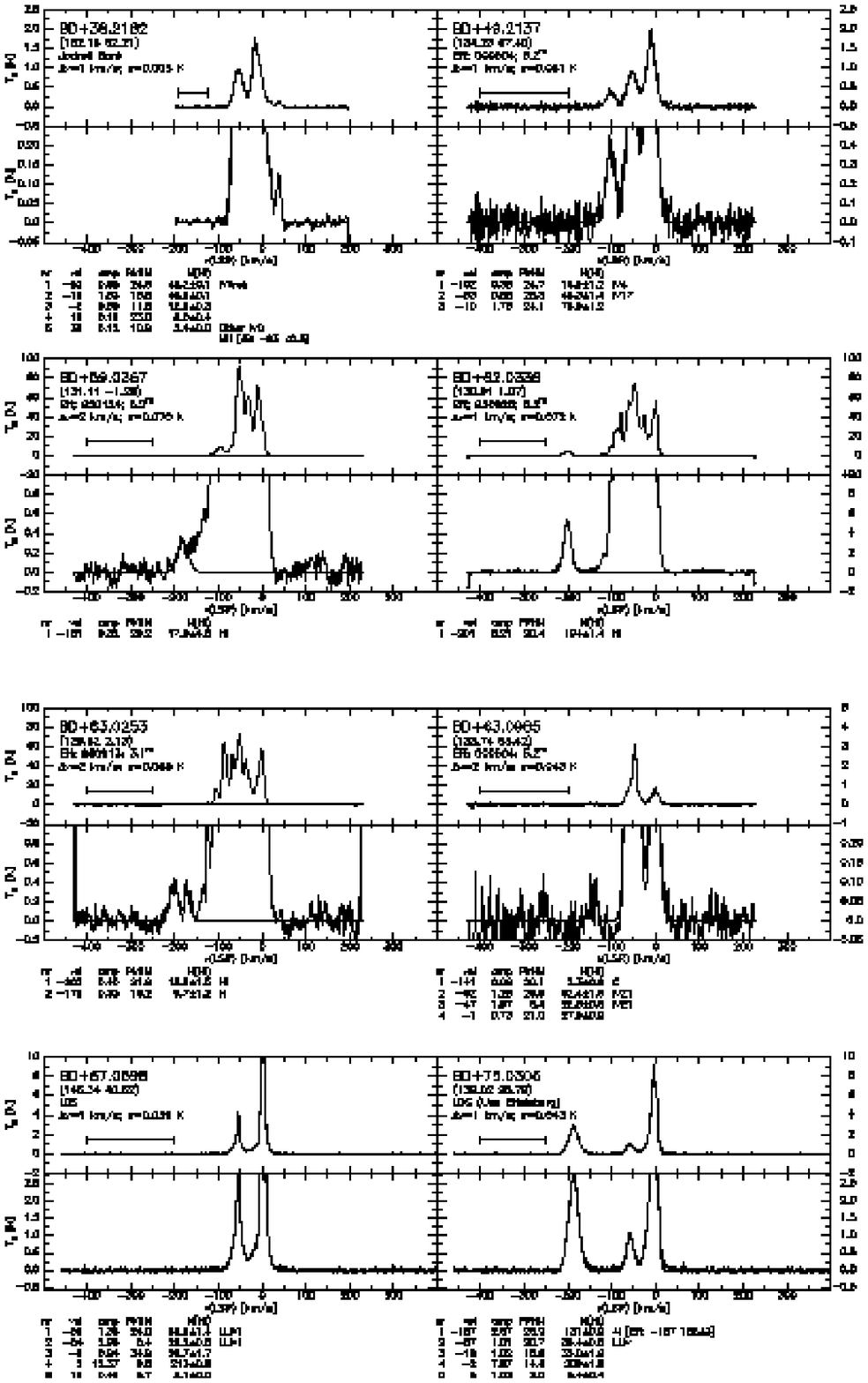}{-45}{-45}{Fig. 1e} %AASPP
\InsertPage{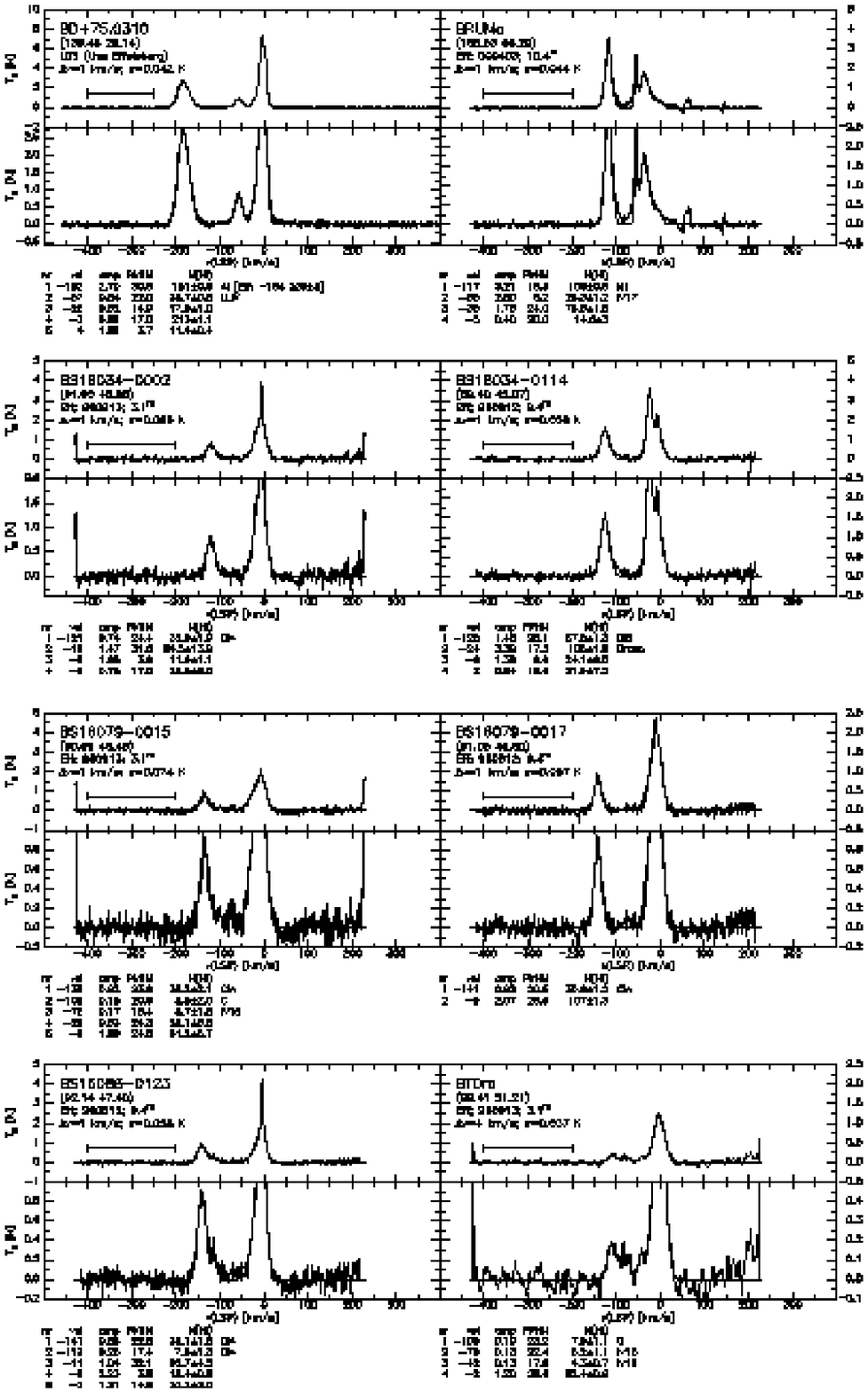}{-45}{-45}{Fig. 1f} %AASPP
\InsertPage{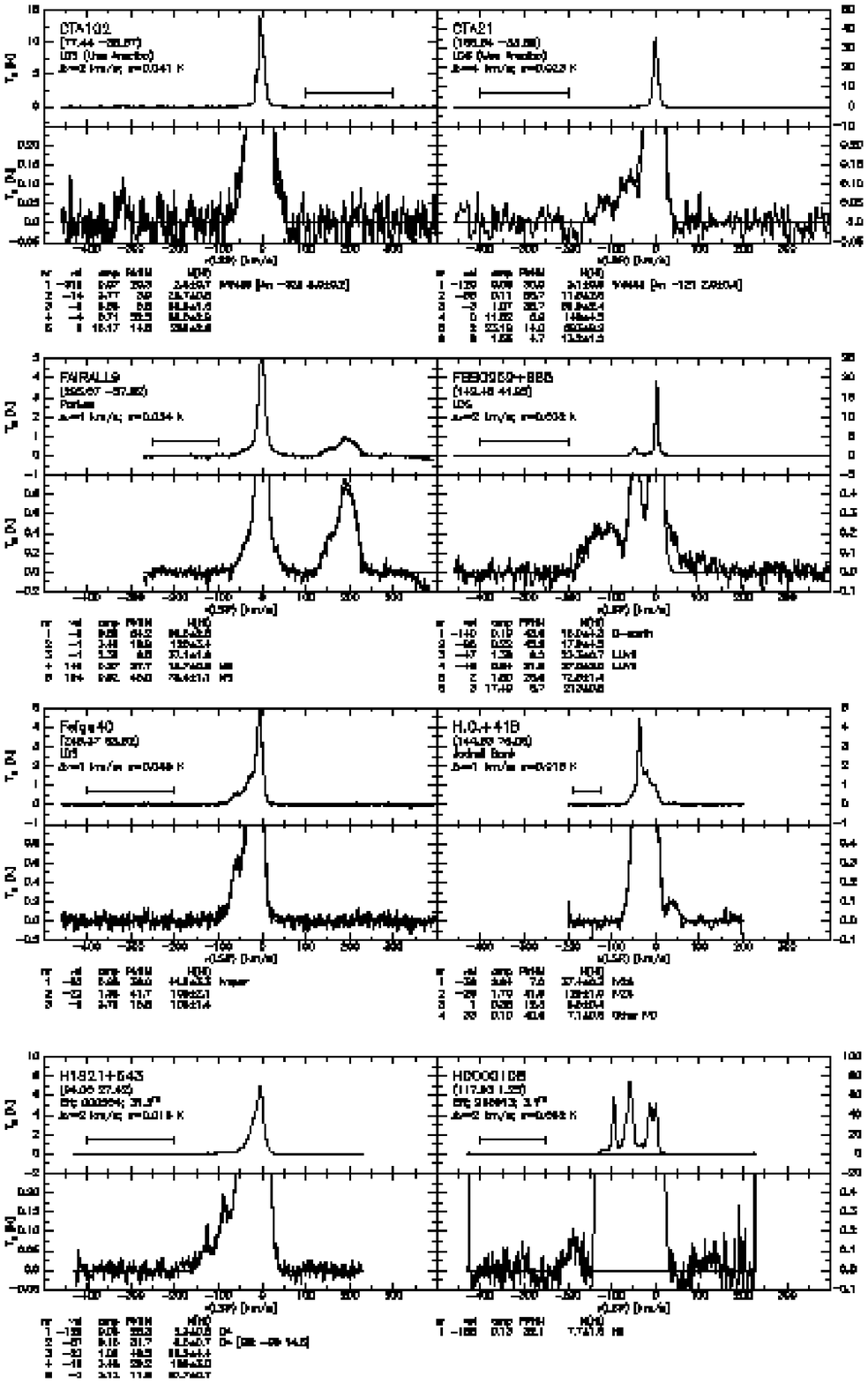}{-45}{-45}{Fig. 1g} %AASPP
\InsertPage{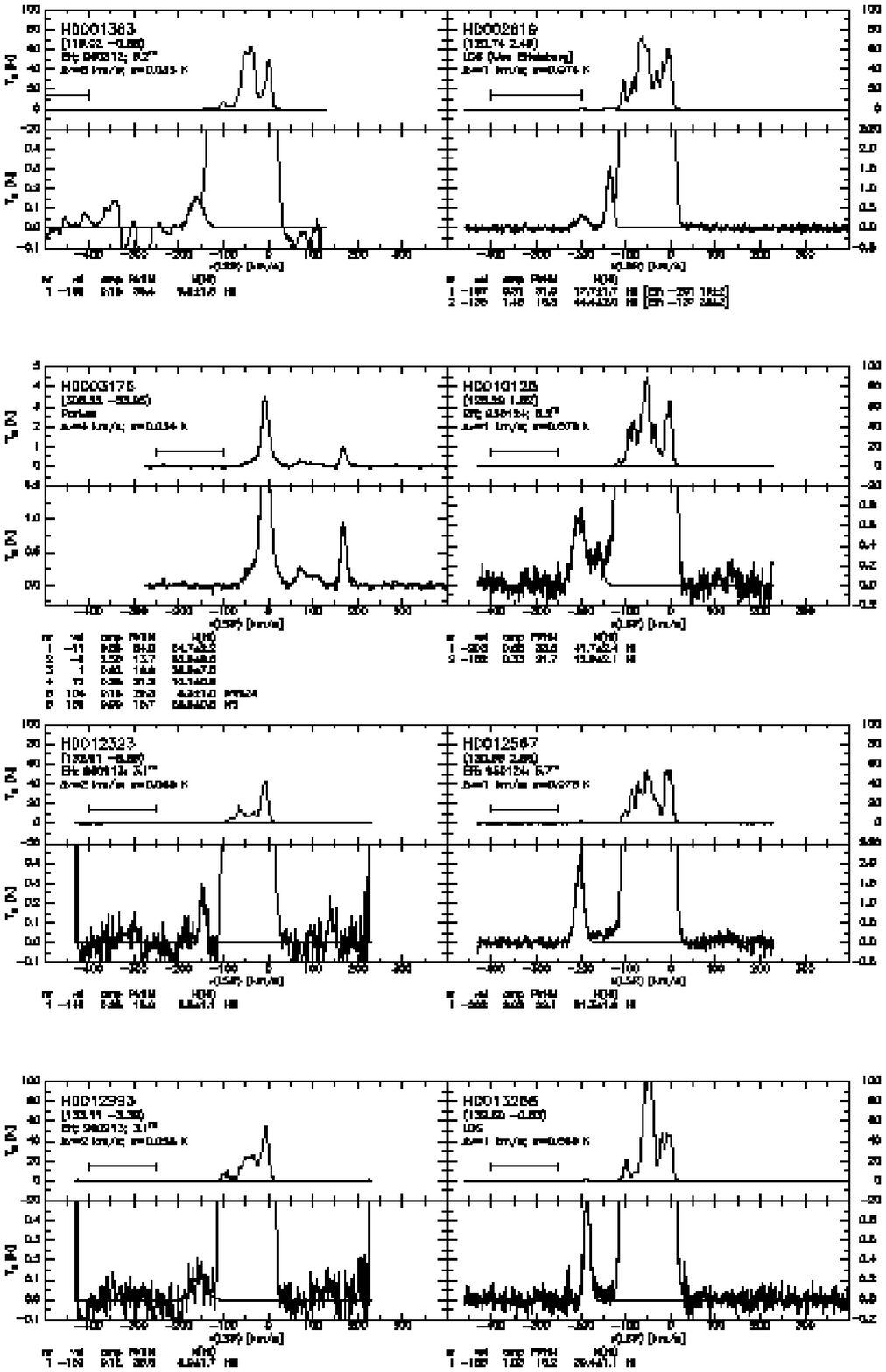}{-45}{-45}{Fig. 1h} %AASPP
\InsertPage{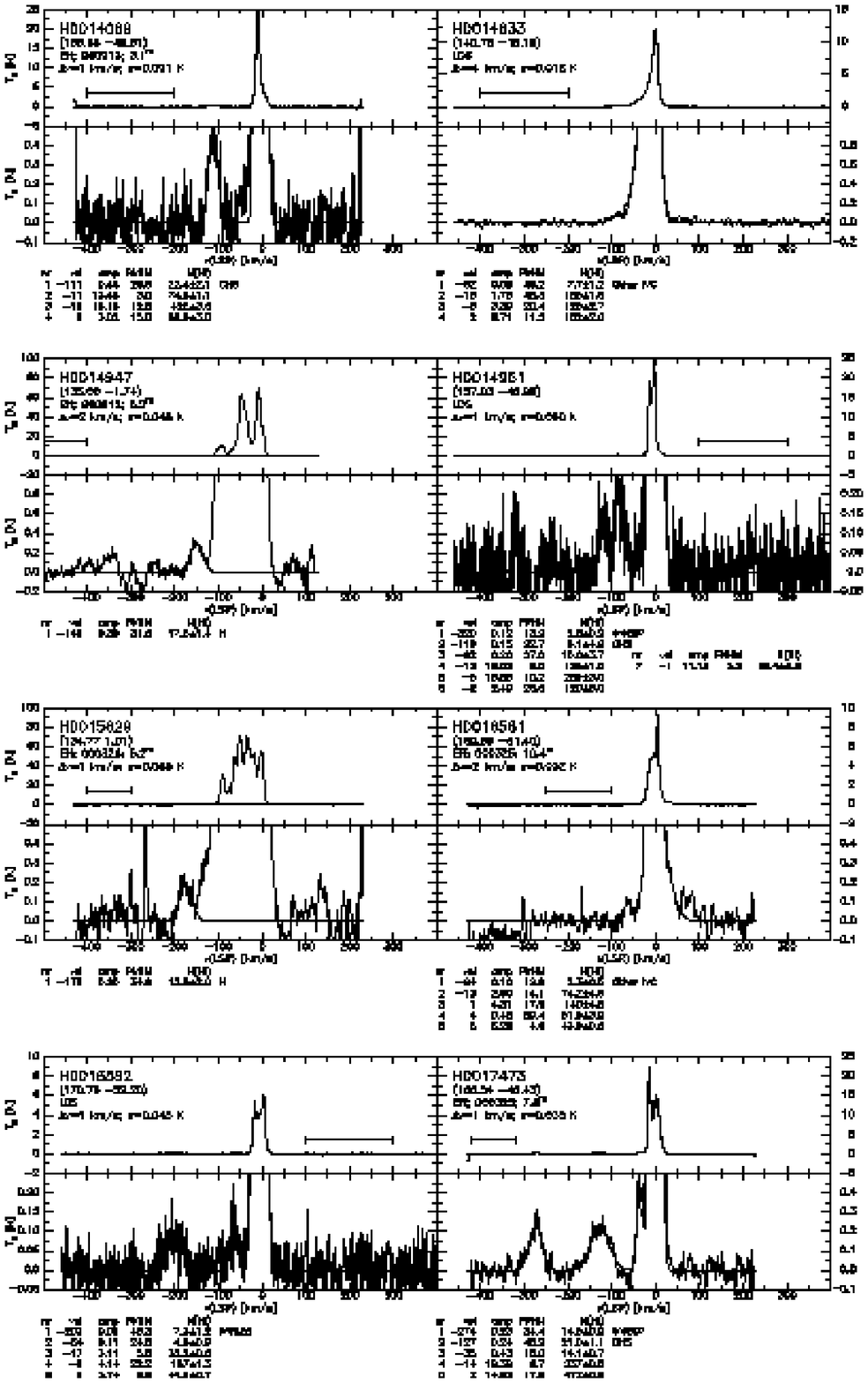}{-45}{-45}{Fig. 1i} %AASPP
\InsertPage{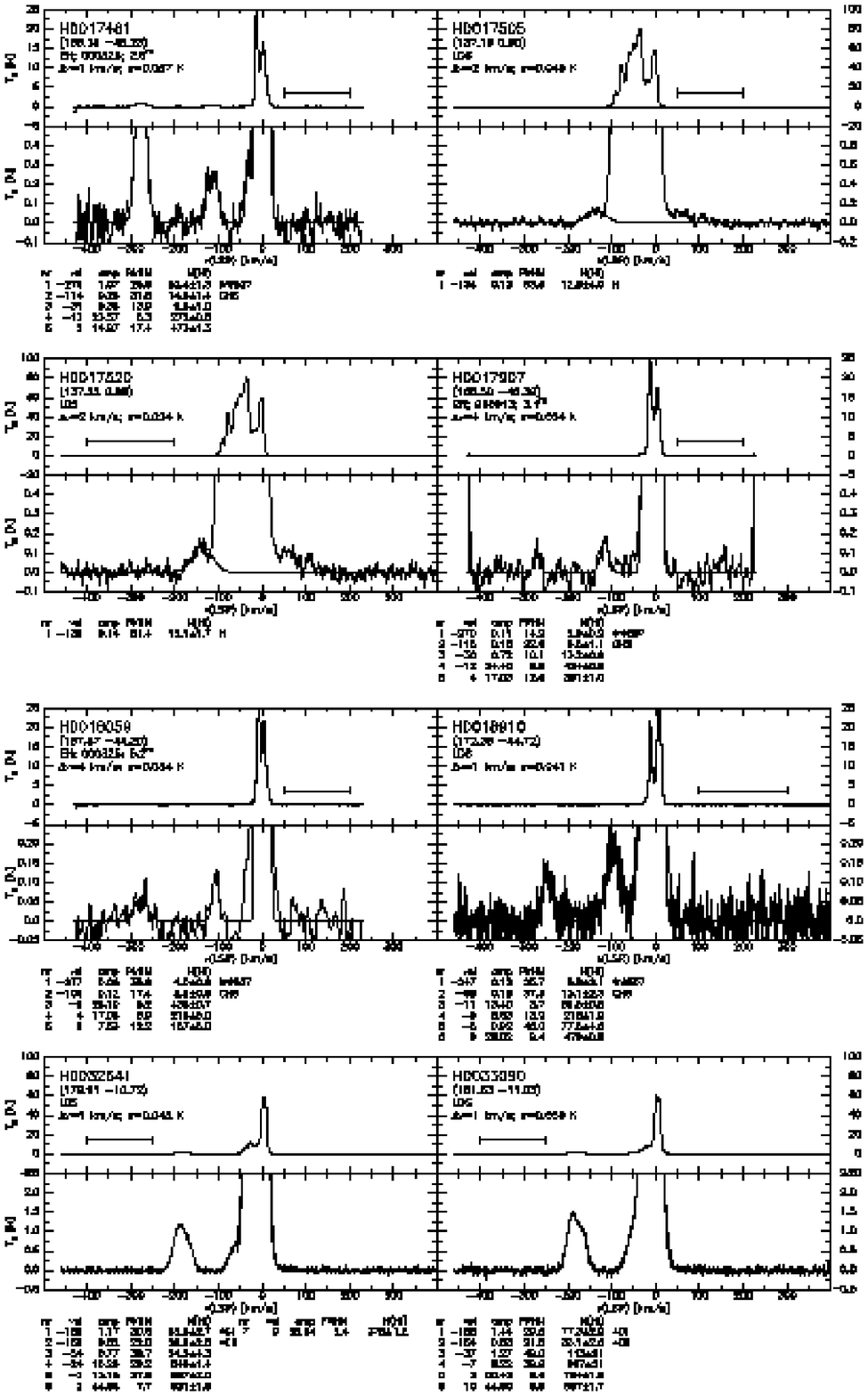}{-45}{-45}{Fig. 1j} %AASPP
\InsertPage{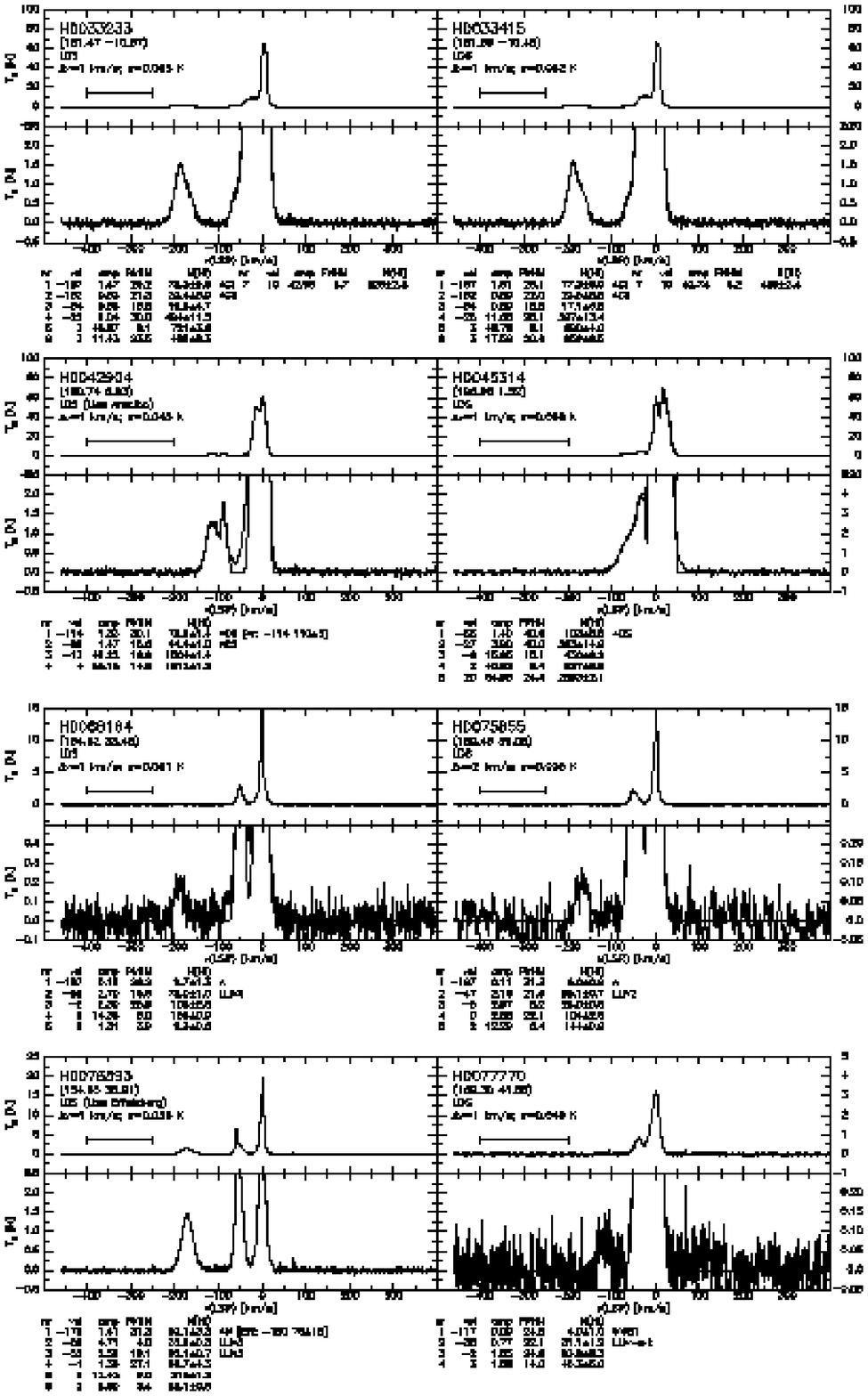}{-45}{-45}{Fig. 1k} %AASPP
\InsertPage{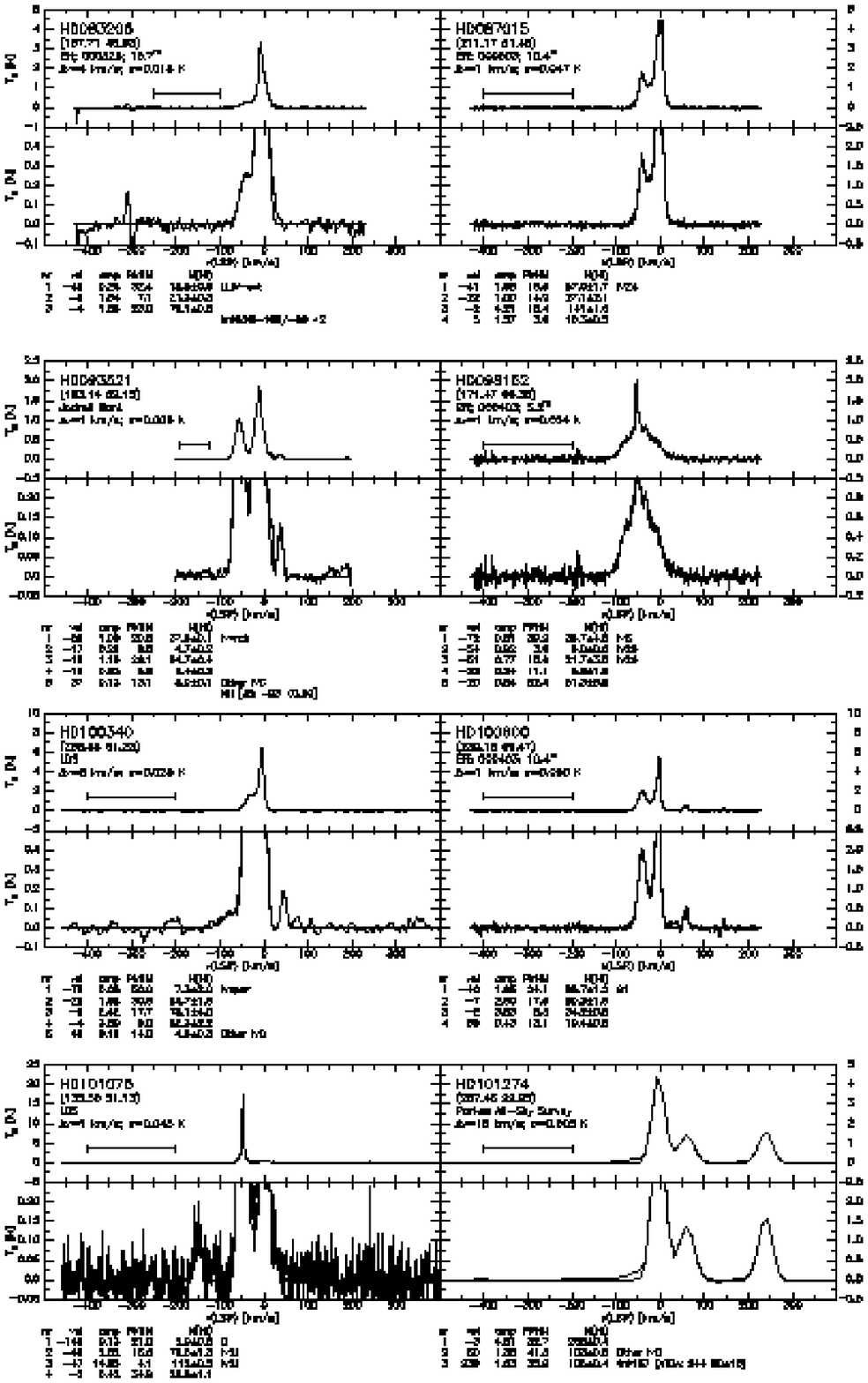}{-45}{-45}{Fig. 1l} %AASPP
\InsertPage{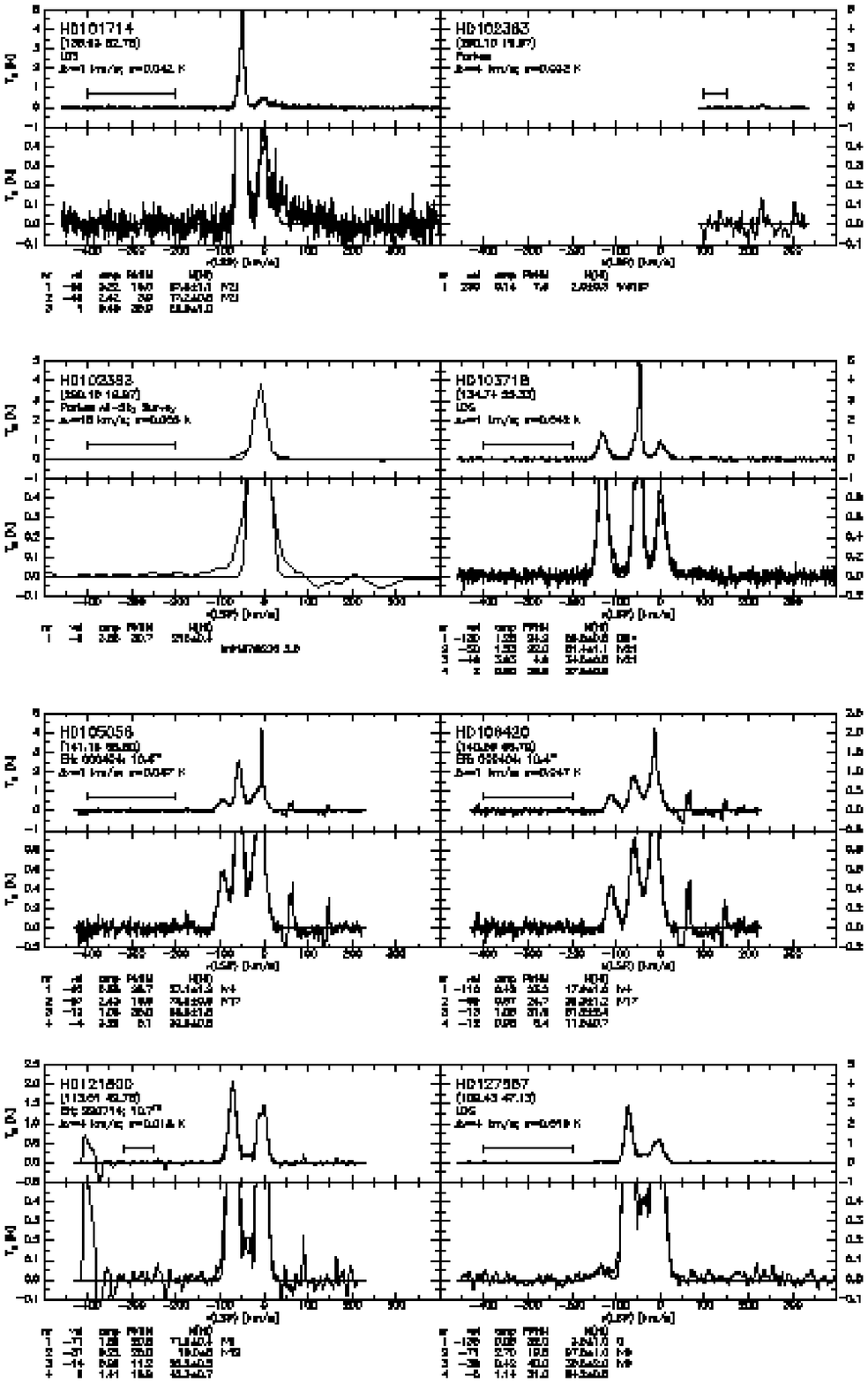}{-45}{-45}{Fig. 1m} %AASPP
\InsertPage{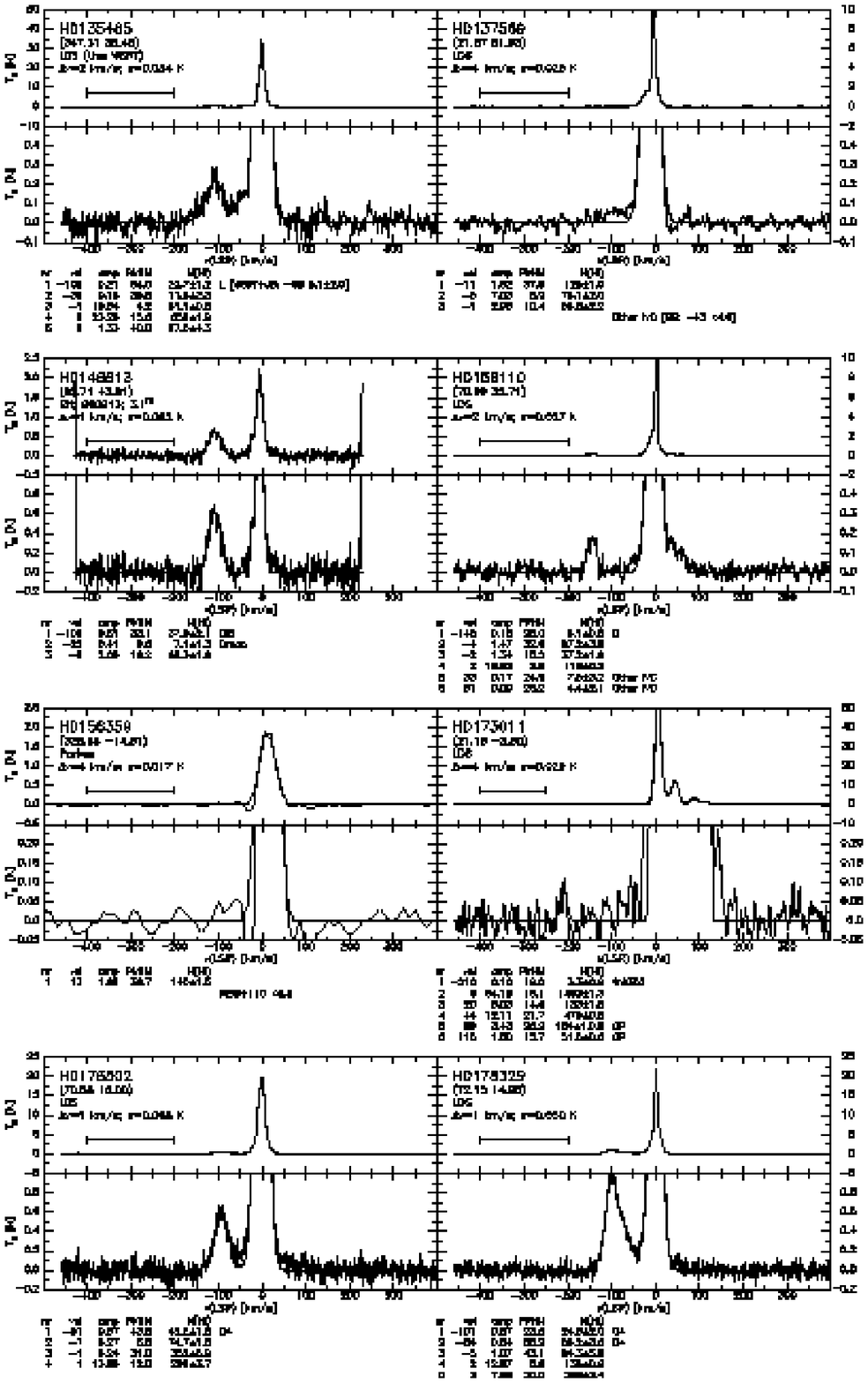}{-45}{-45}{Fig. 1n} %AASPP
\InsertPage{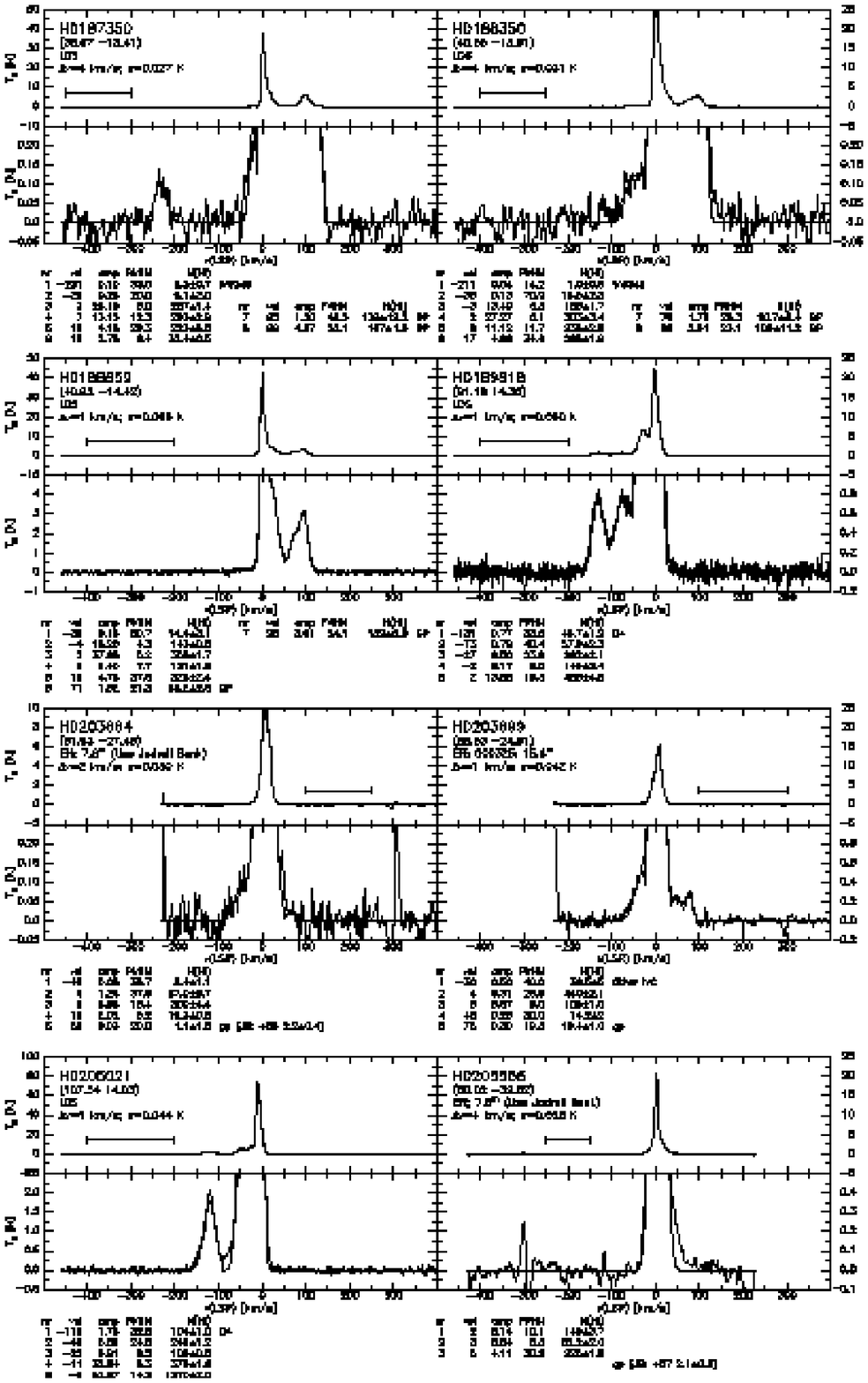}{-45}{-45}{Fig. 1o} %AASPP
\InsertPage{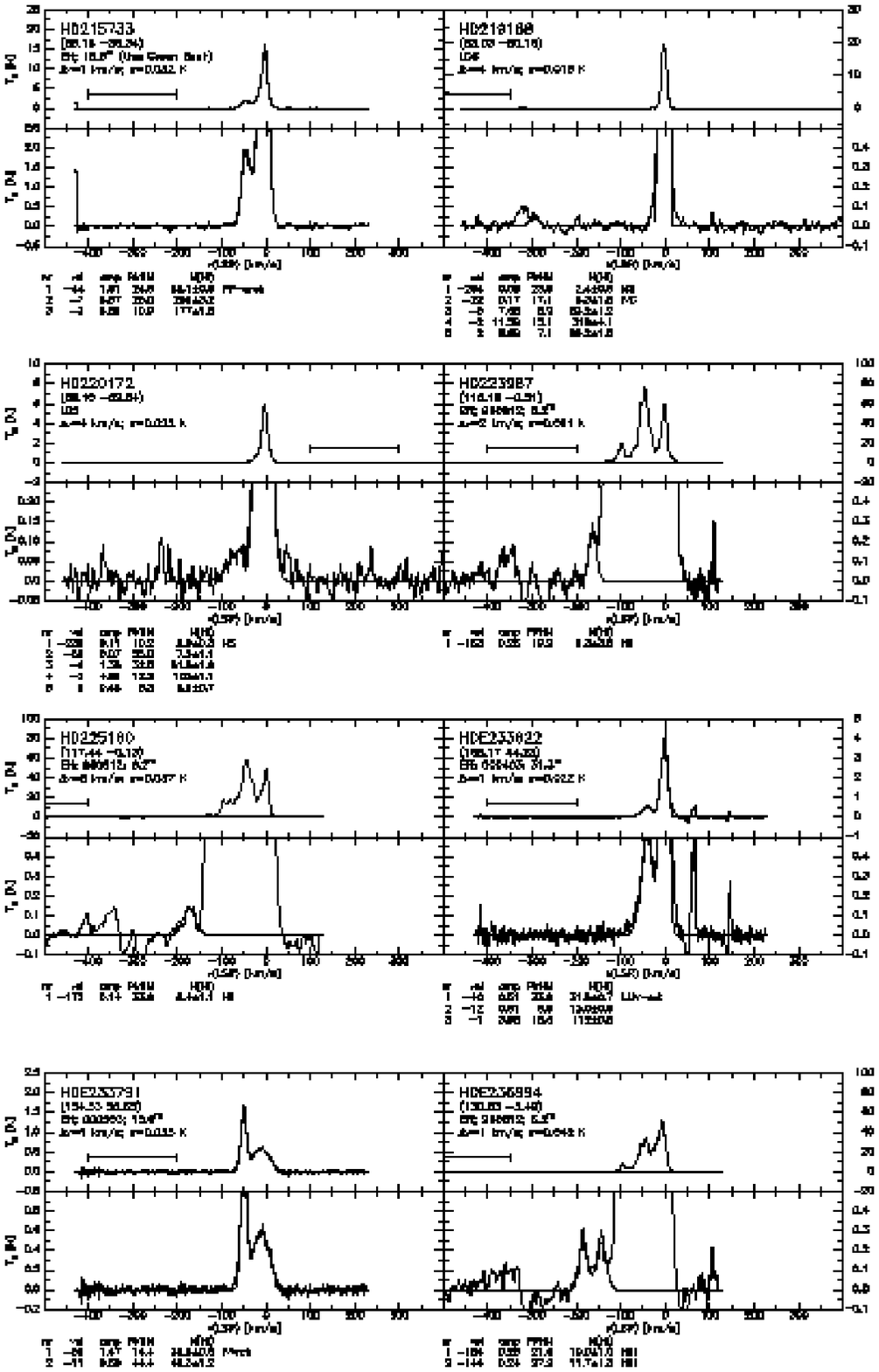}{-45}{-45}{Fig. 1p} %AASPP
\InsertPage{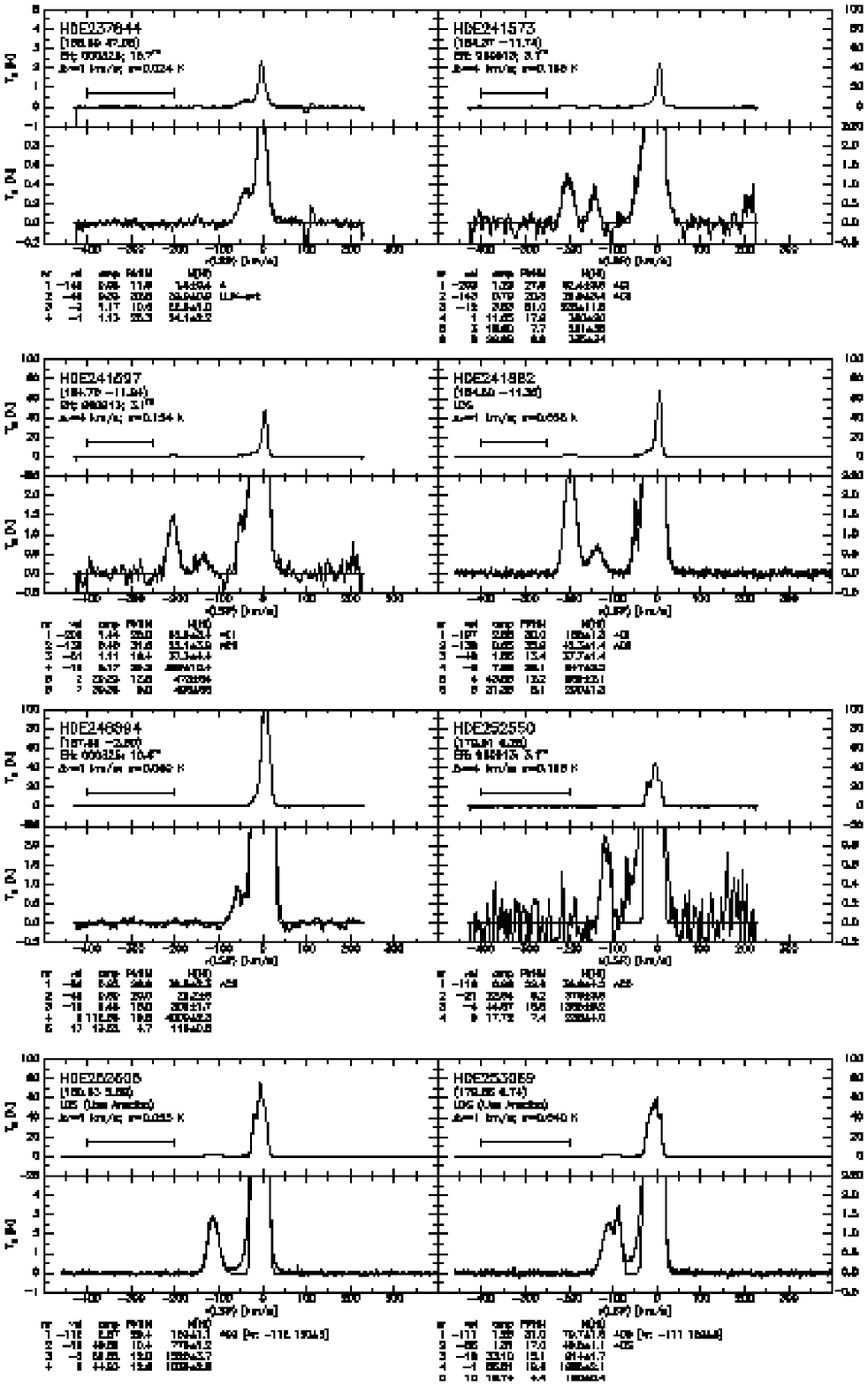}{-45}{-45}{Fig. 1q} %AASPP
\InsertPage{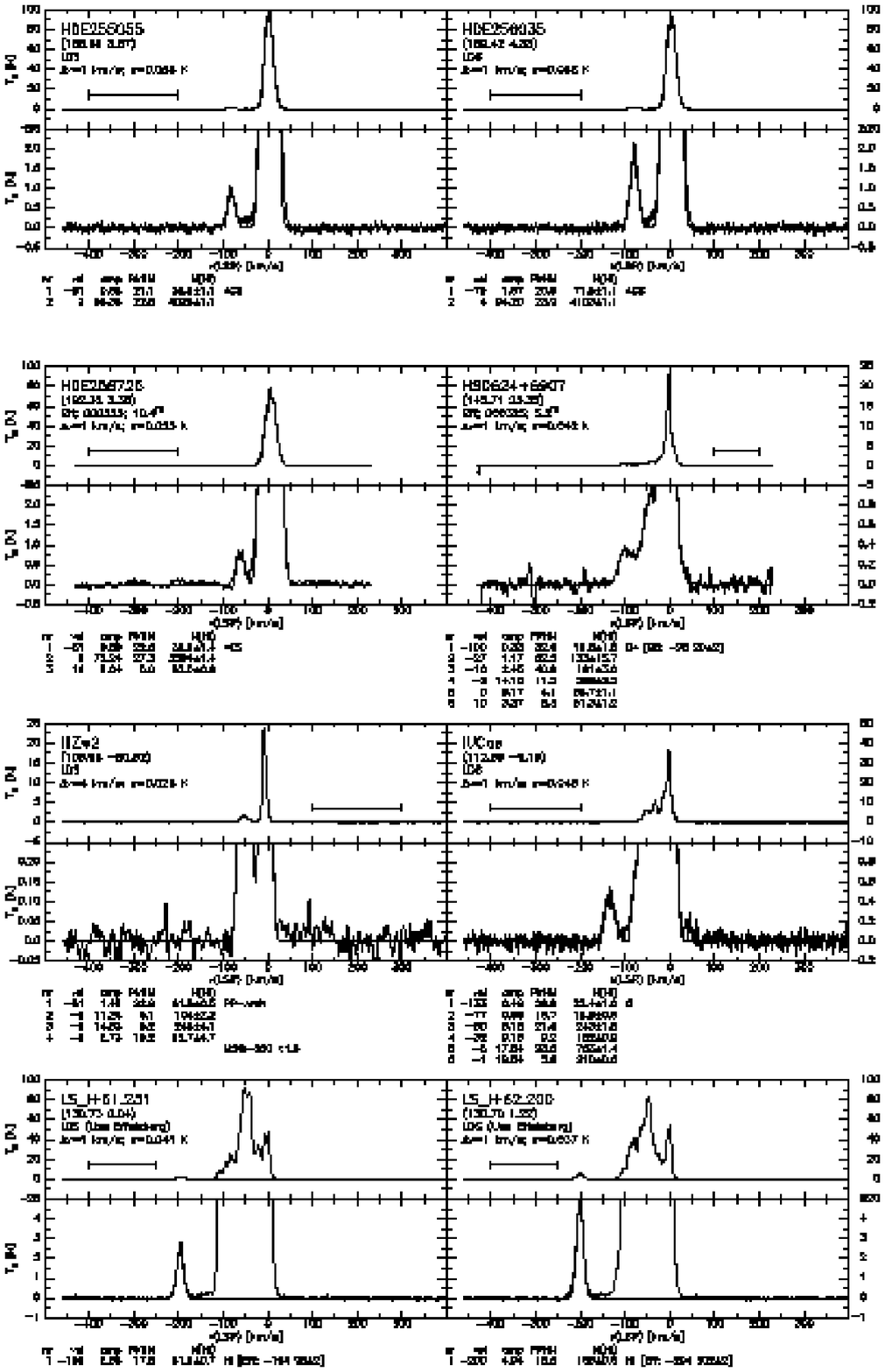}{-45}{-45}{Fig. 1r} %AASPP
\InsertPage{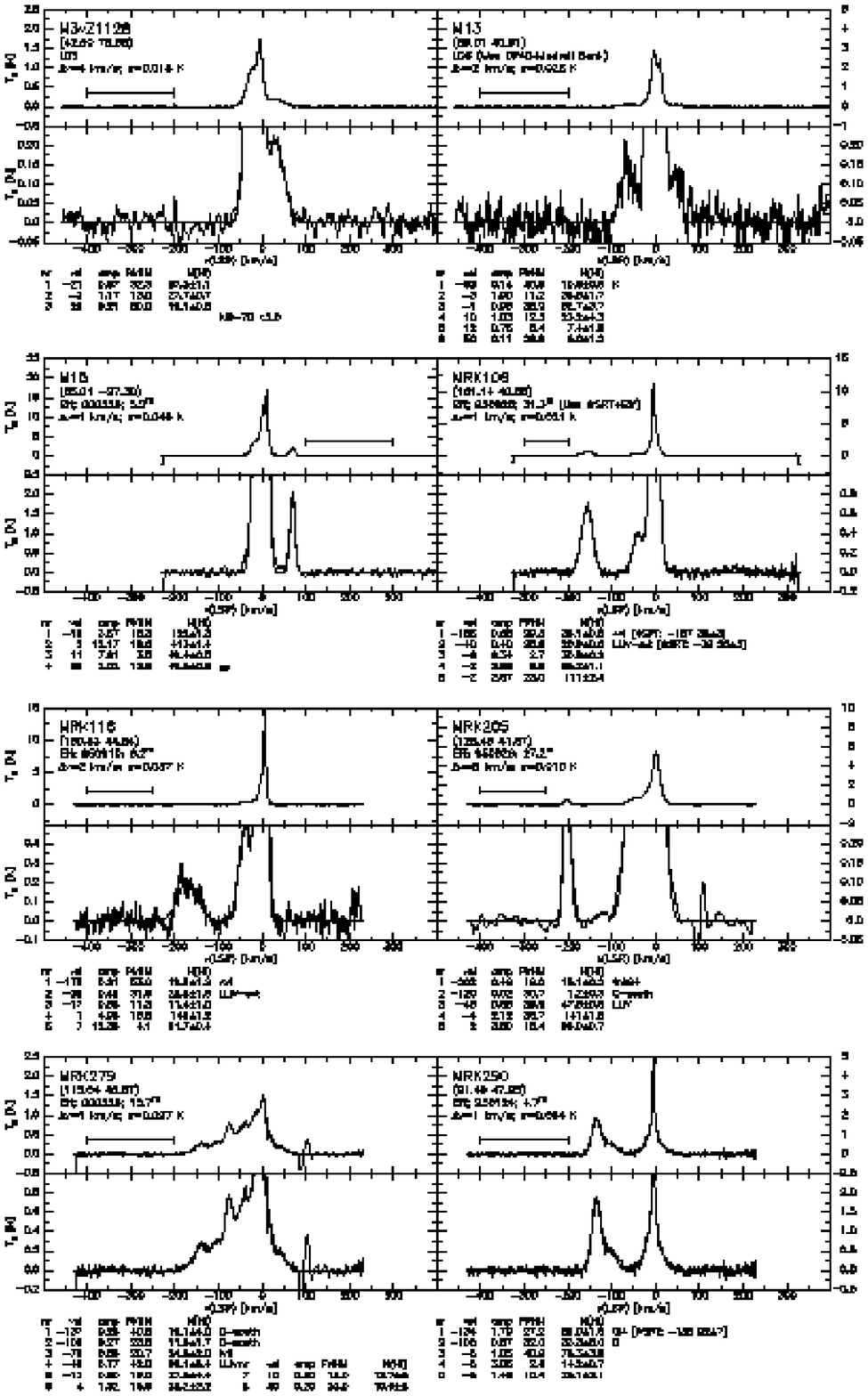}{-45}{-45}{Fig. 1s} %AASPP
\InsertPage{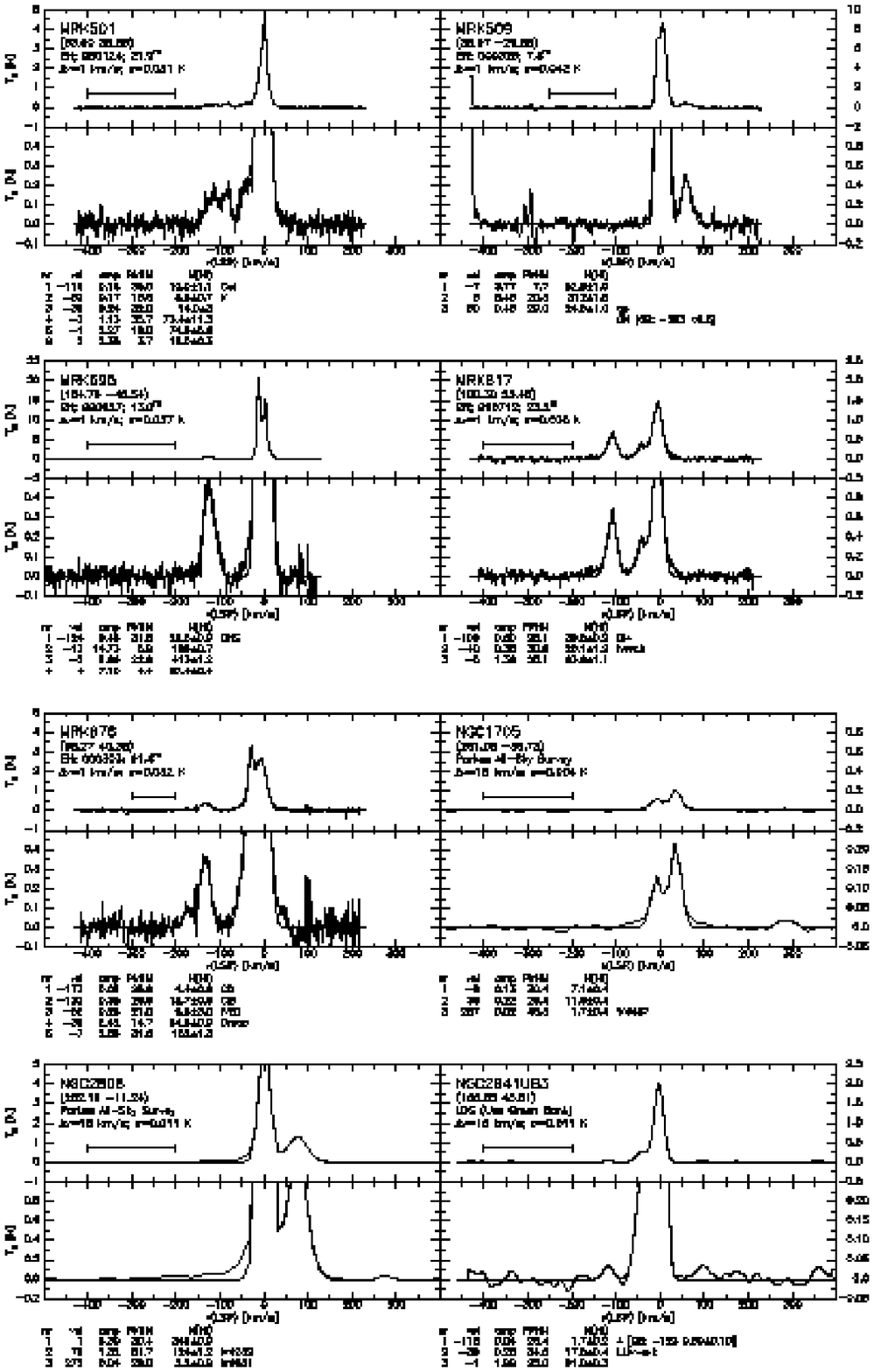}{-45}{-45}{Fig. 1t} %AASPP
\InsertPage{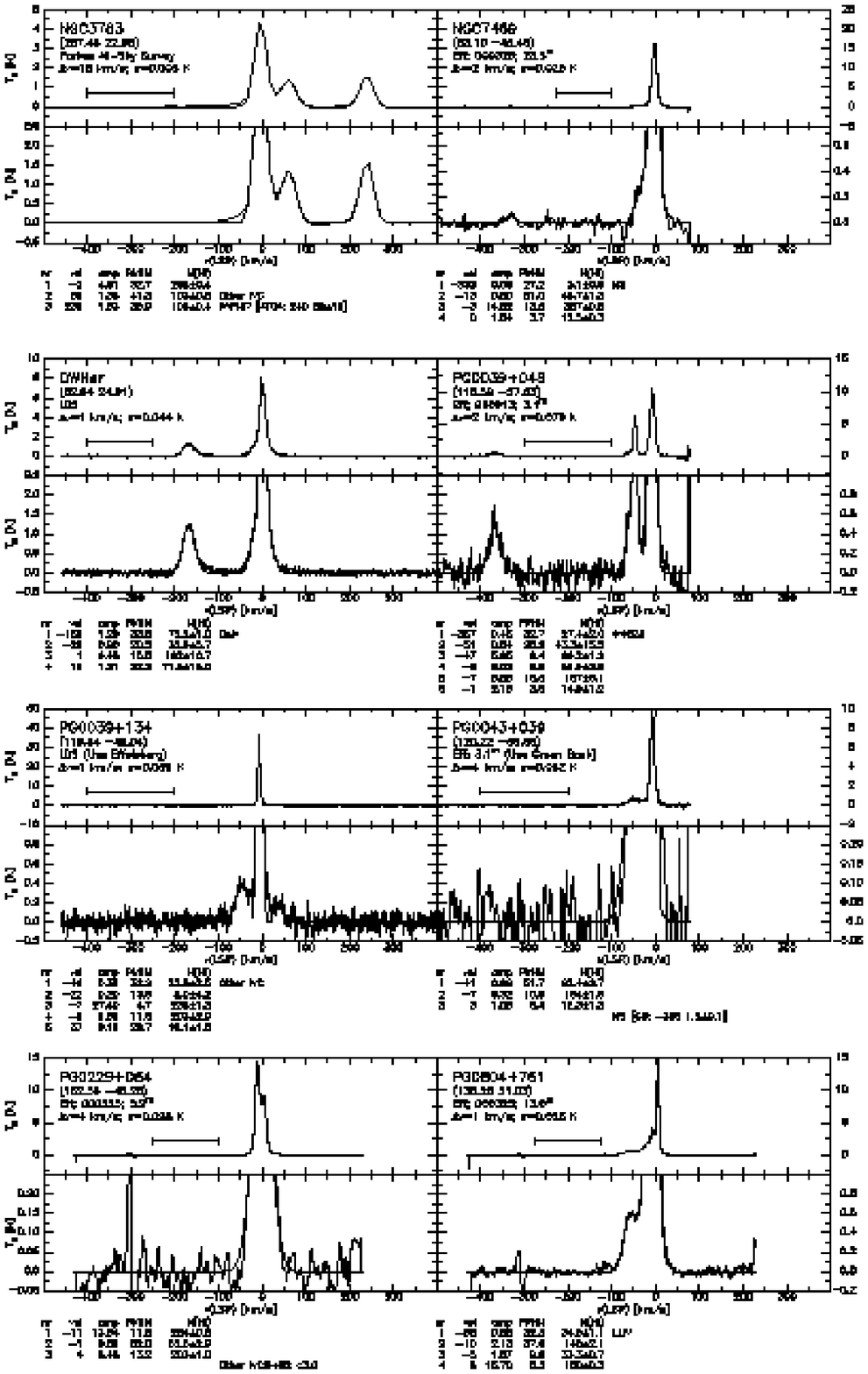}{-45}{-45}{Fig. 1u} %AASPP
\InsertPage{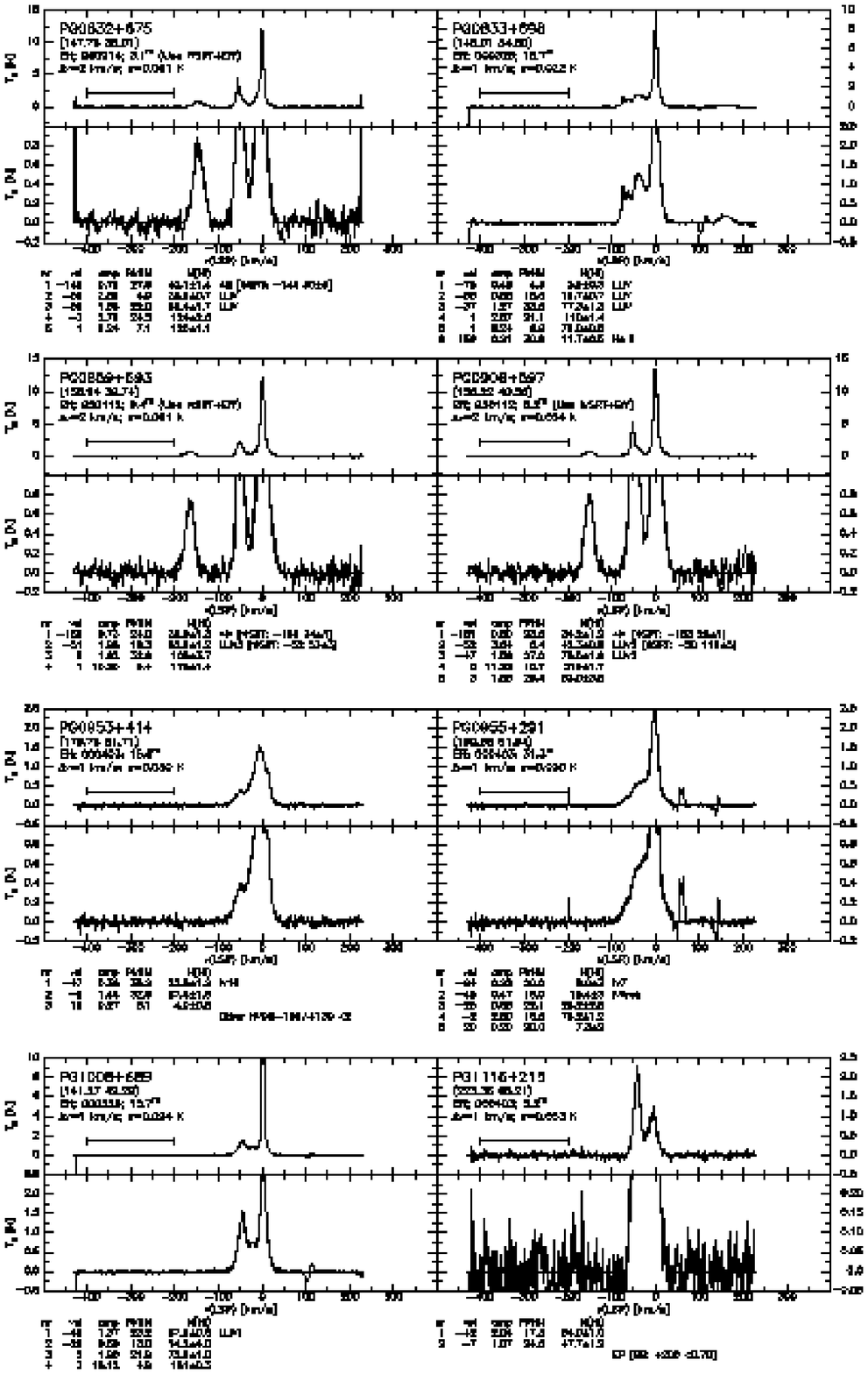}{-45}{-45}{Fig. 1v} %AASPP
\InsertPage{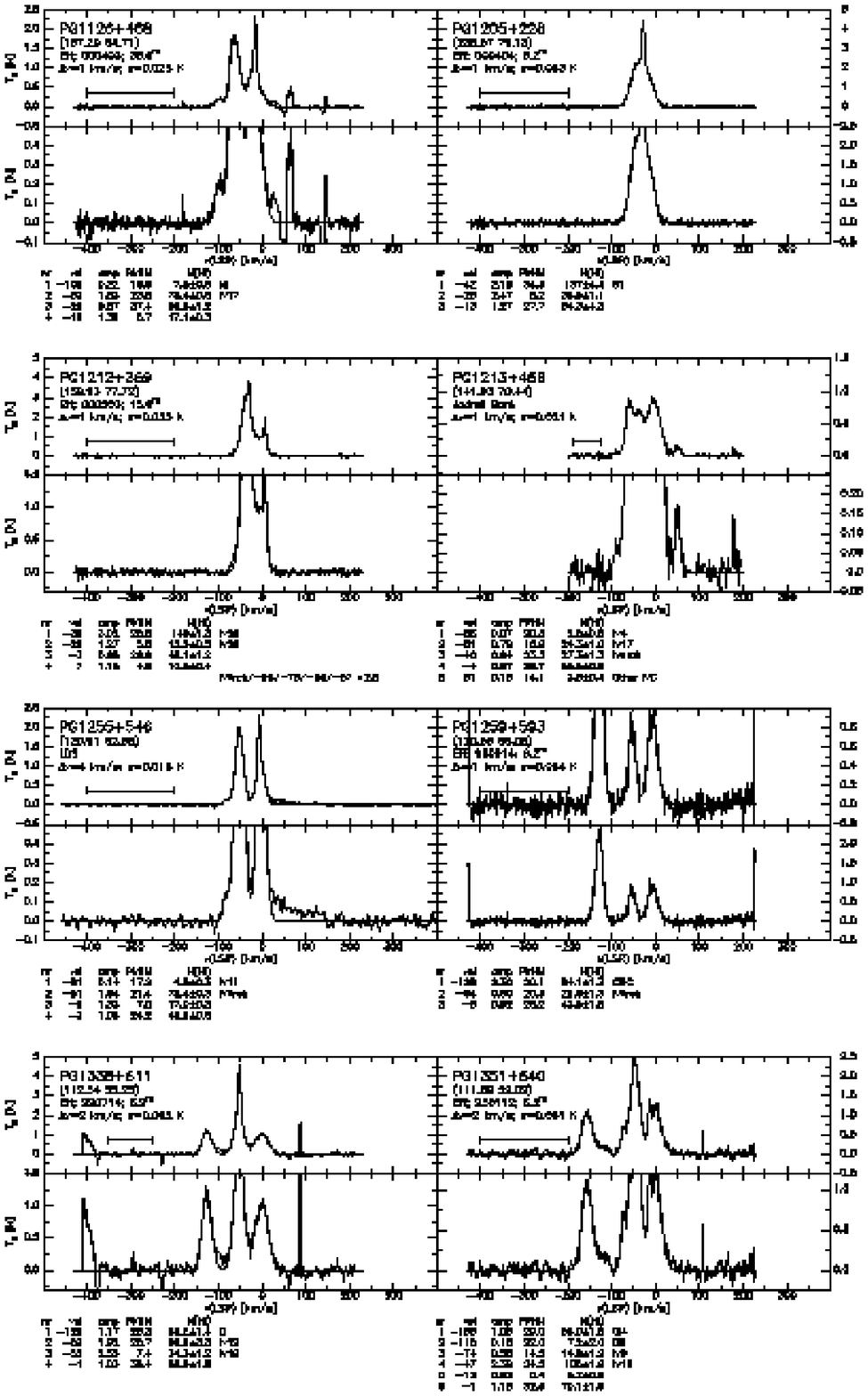}{-45}{-45}{Fig. 1w} %AASPP
\InsertPage{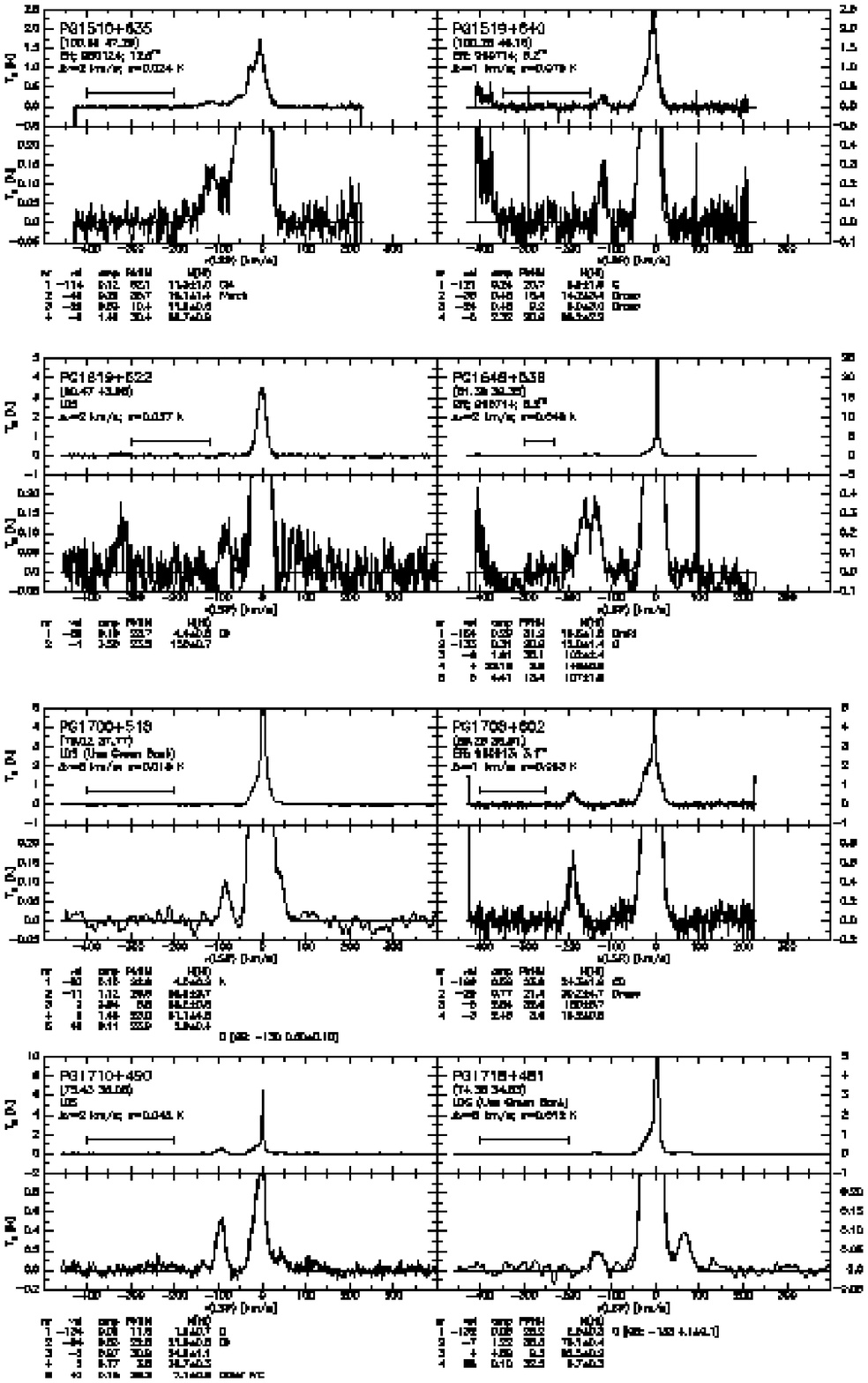}{-45}{-45}{Fig. 1x} %AASPP
\InsertPage{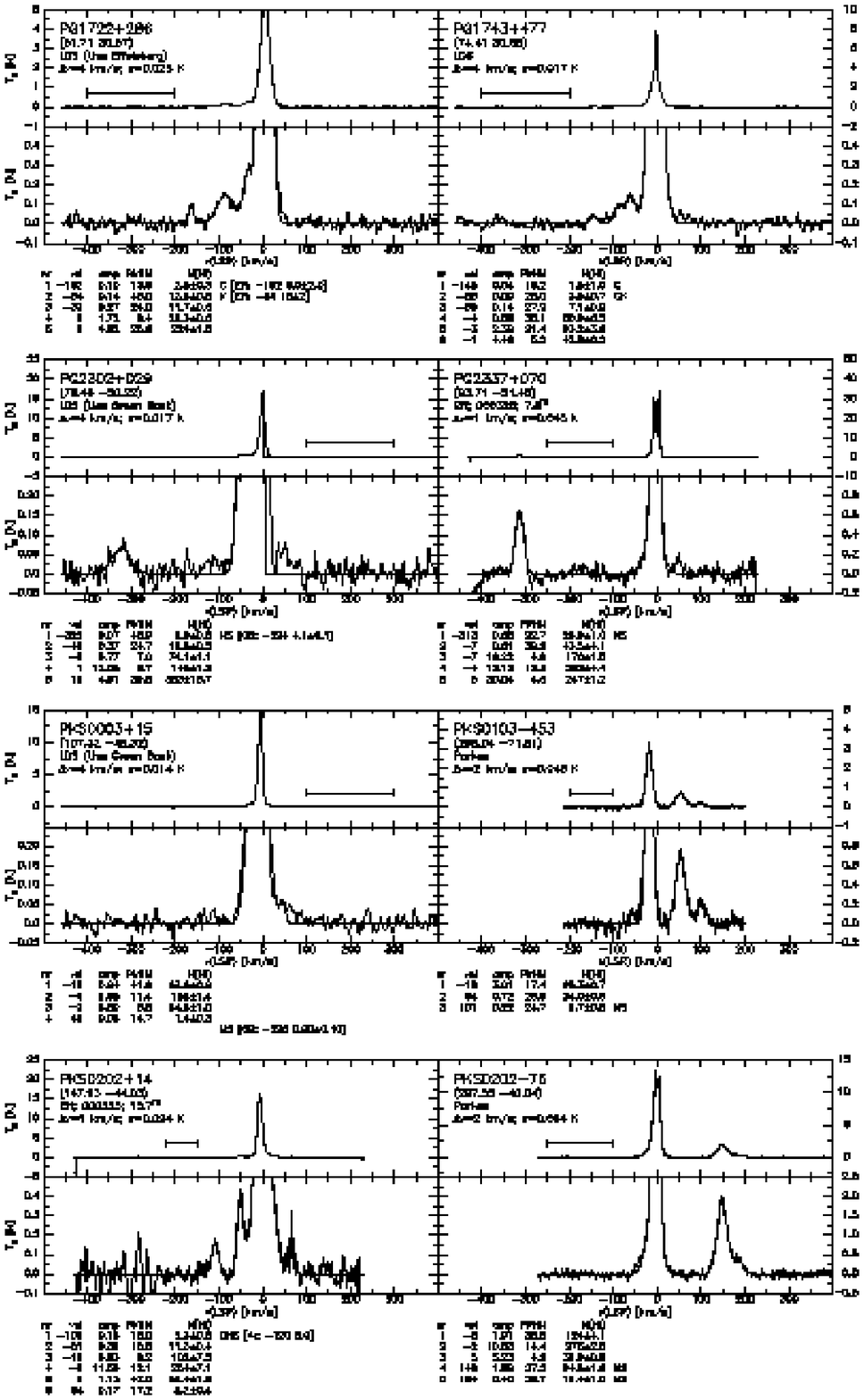}{-45}{-45}{Fig. 1y} %AASPP
\InsertPage{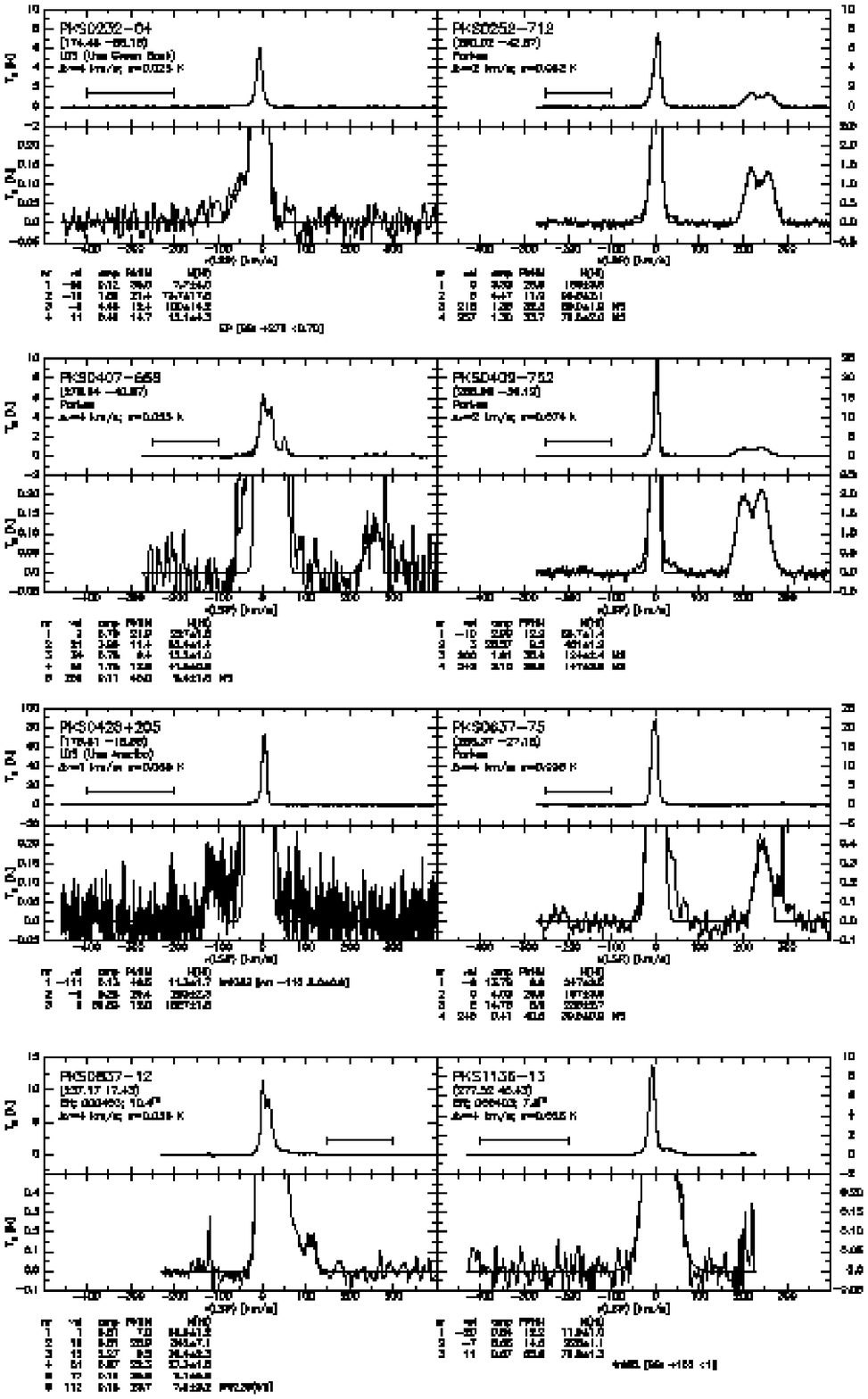}{-45}{-45}{Fig. 1z} %AASPP
\InsertPage{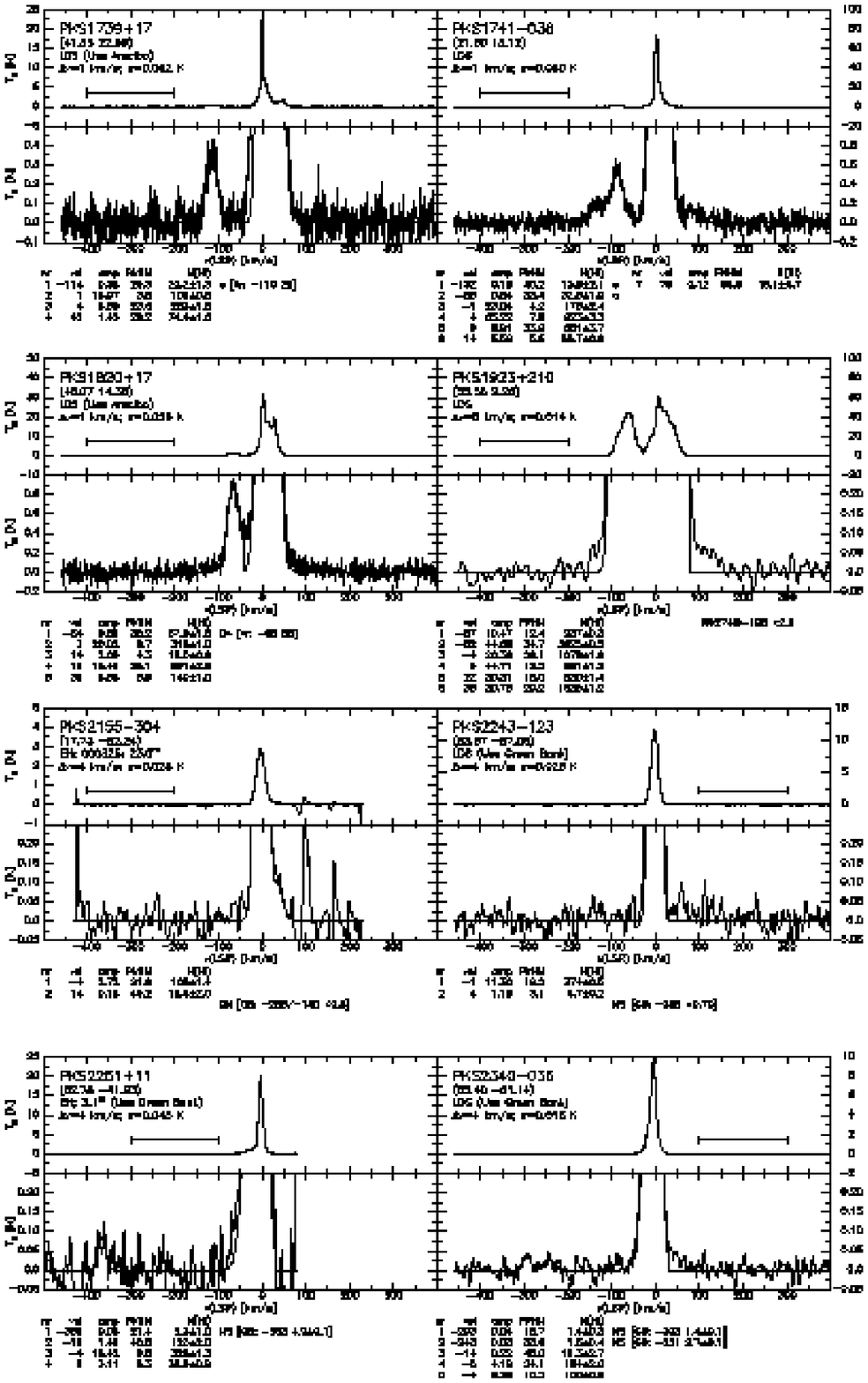}{-45}{-45}{Fig. 1aa} %AASPP
\InsertPage{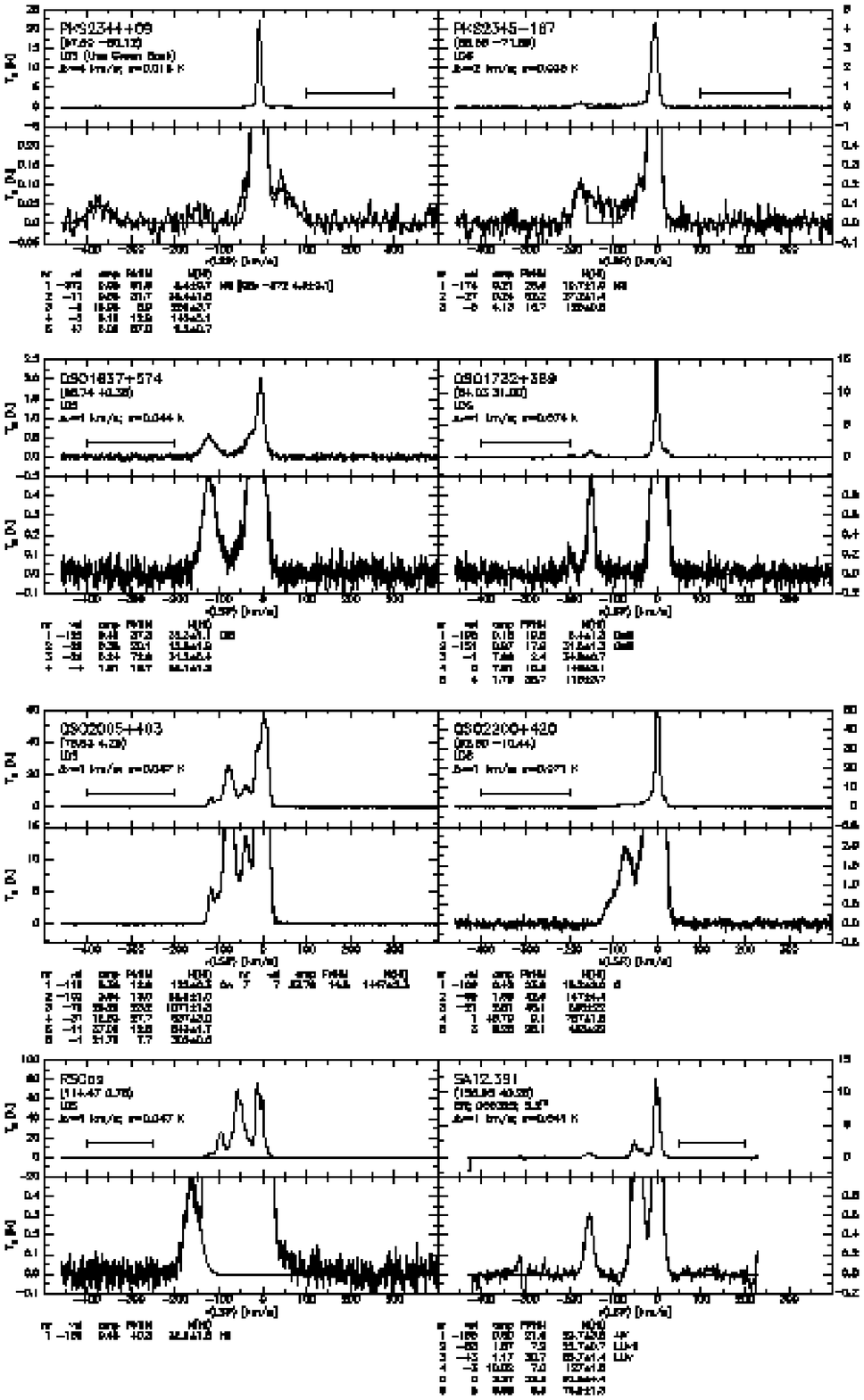}{-45}{-45}{Fig. 1bb} %AASPP
\InsertPage{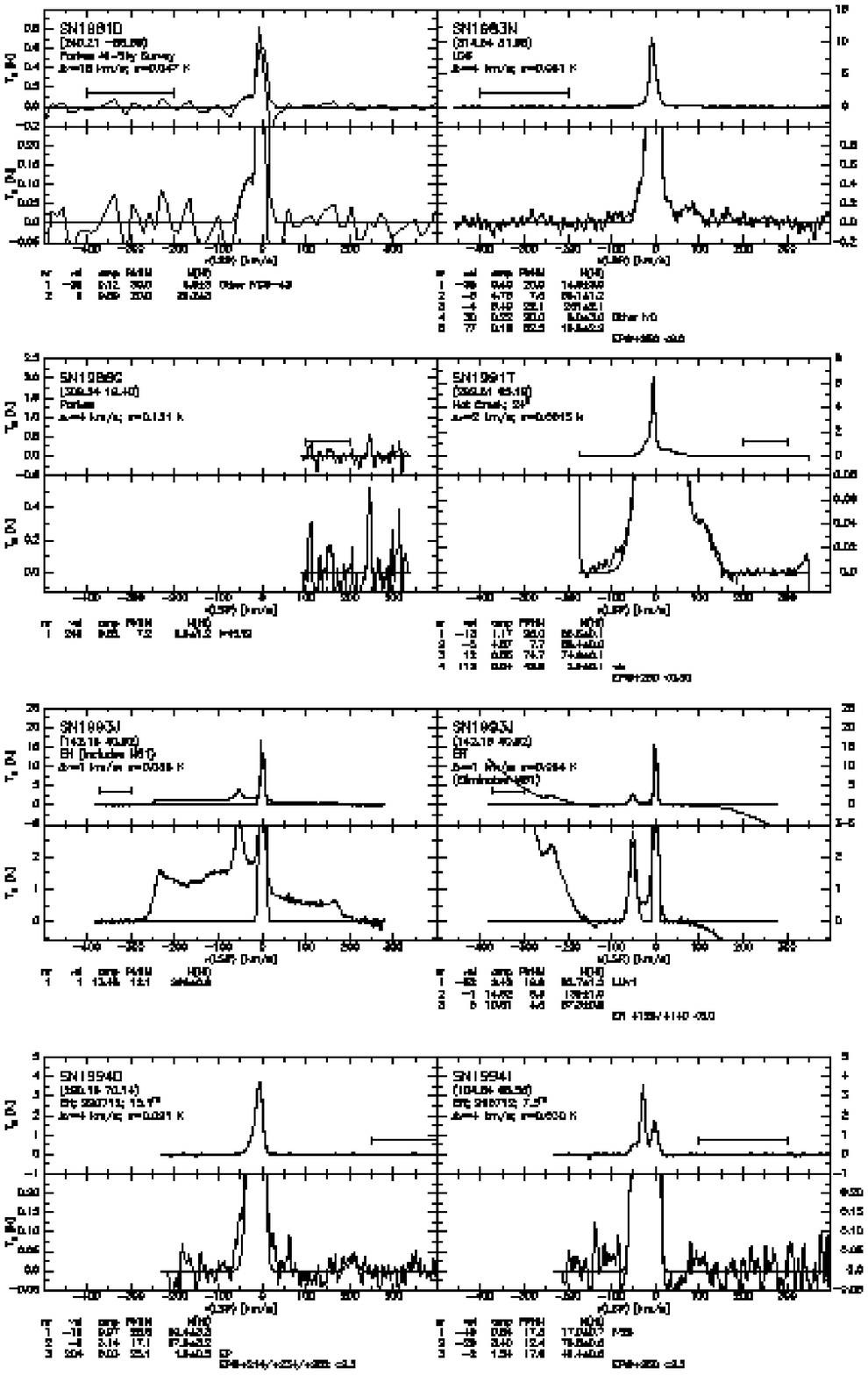}{-45}{-45}{Fig. 1cc} %AASPP
\InsertPage{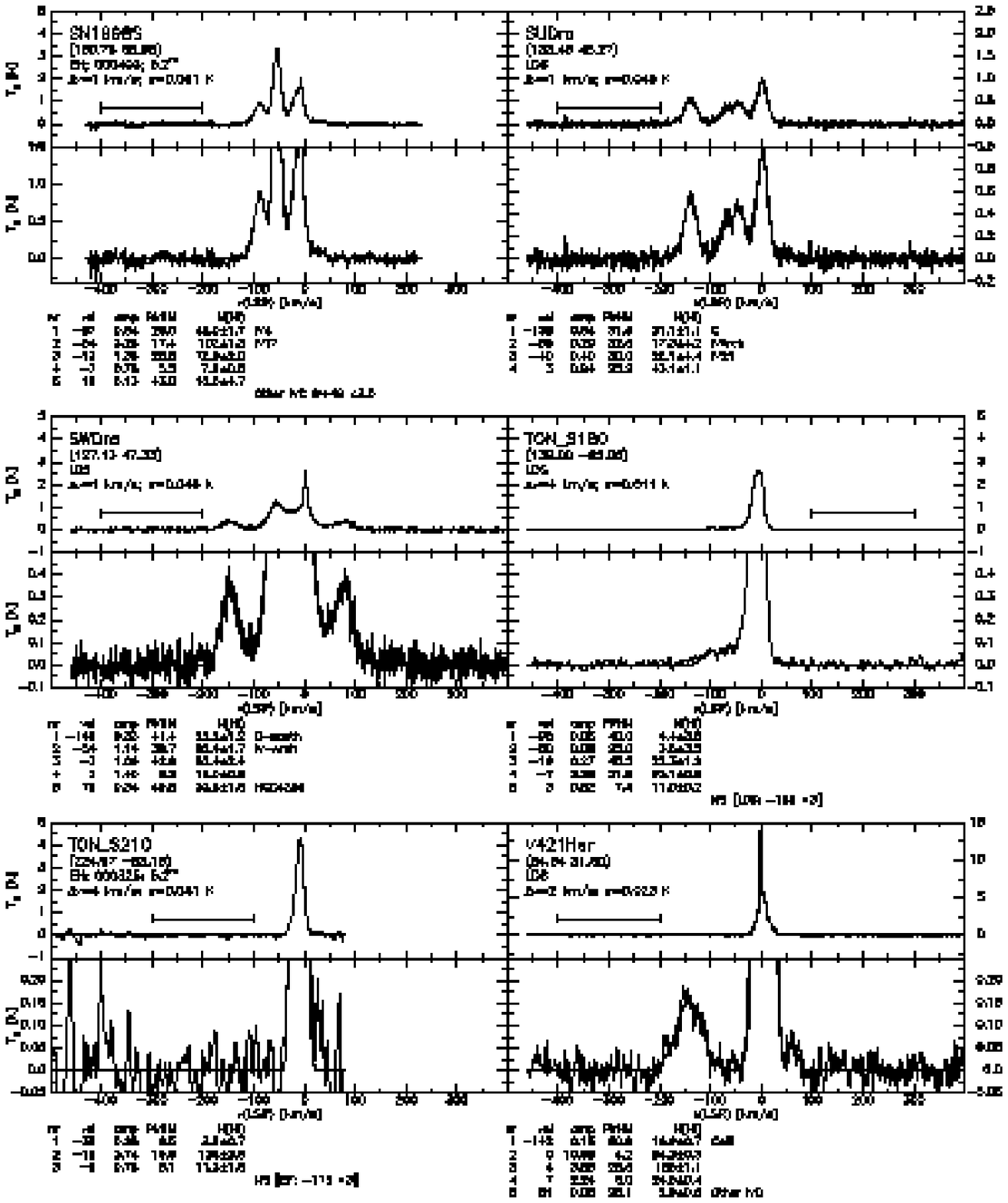}{-45}{-45}{Fig. 1dd} %AASPP

\InsertPage{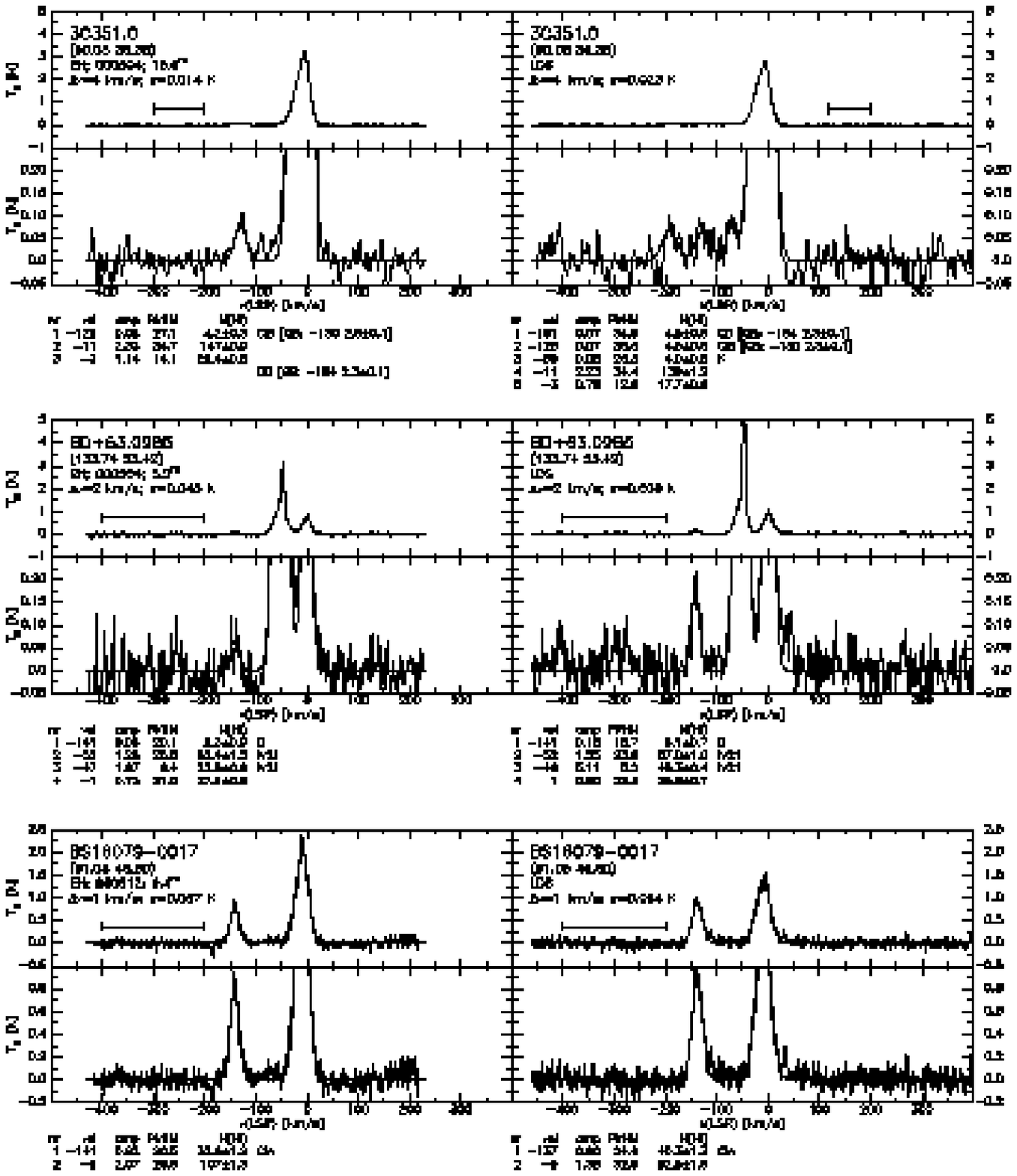}{-45}{-45}{% %AASPP
\fgnumber{2} Eleven examples of the \HI\ profiles seen in the same directions
with Effelsberg (left panels) and with the Dwingeloo telescope (right panels).
The labels are described in detail in Sect.~\Sresults.
}
\InsertPage{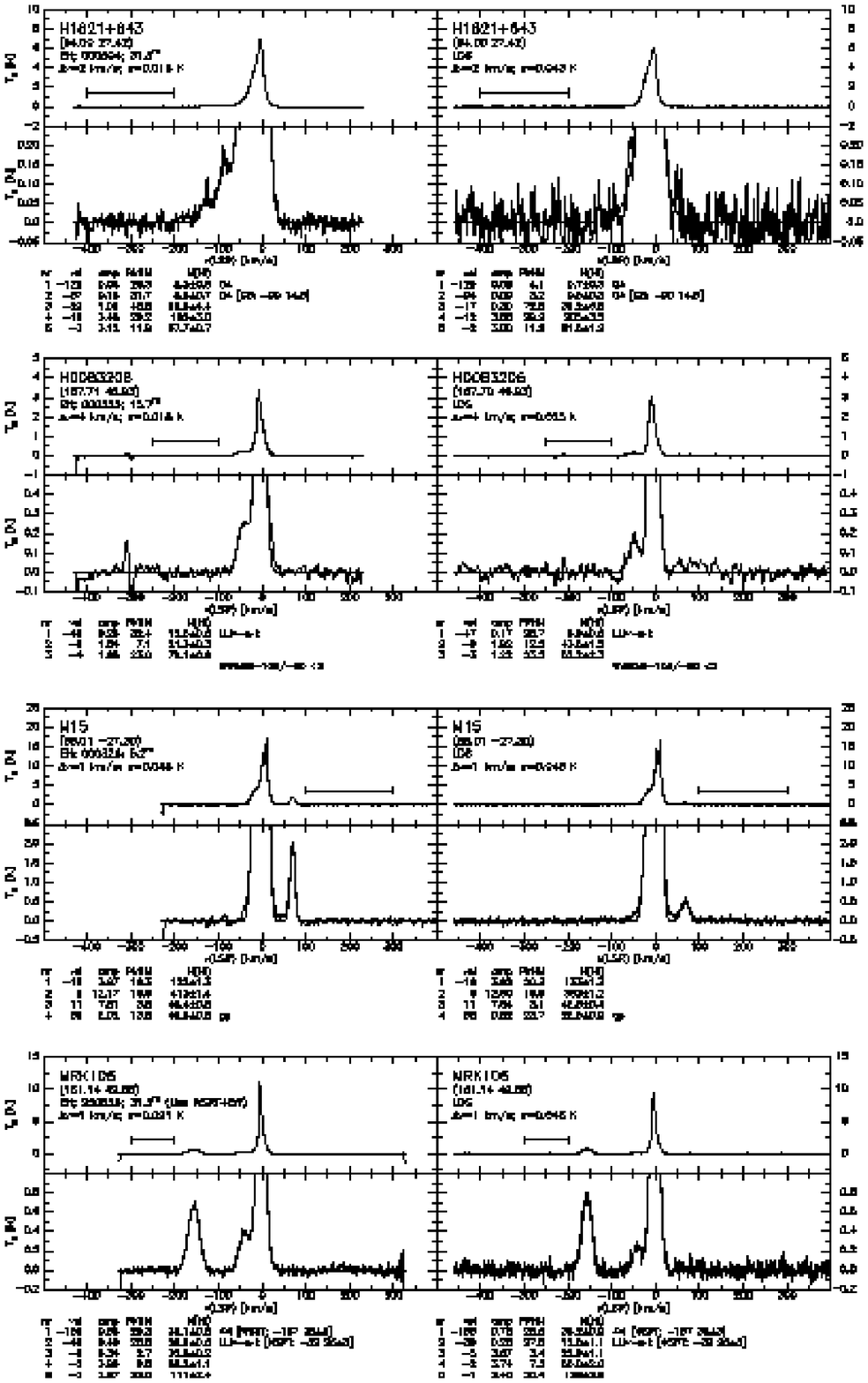}{-45}{-45}{Fig. 2b} %AASPP
\InsertPage{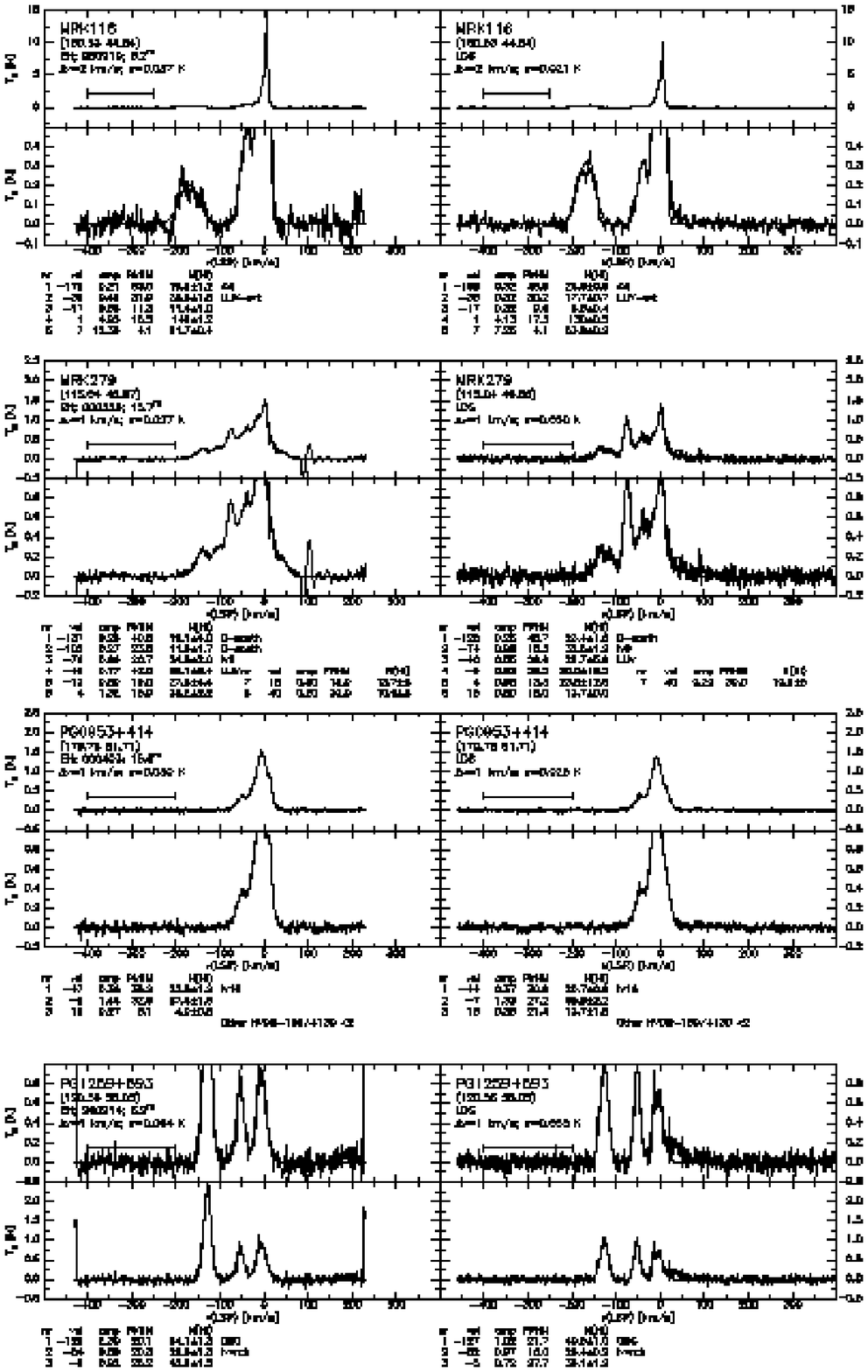}{-45}{-45}{Fig. 2c} %AASPP

\InsertPage{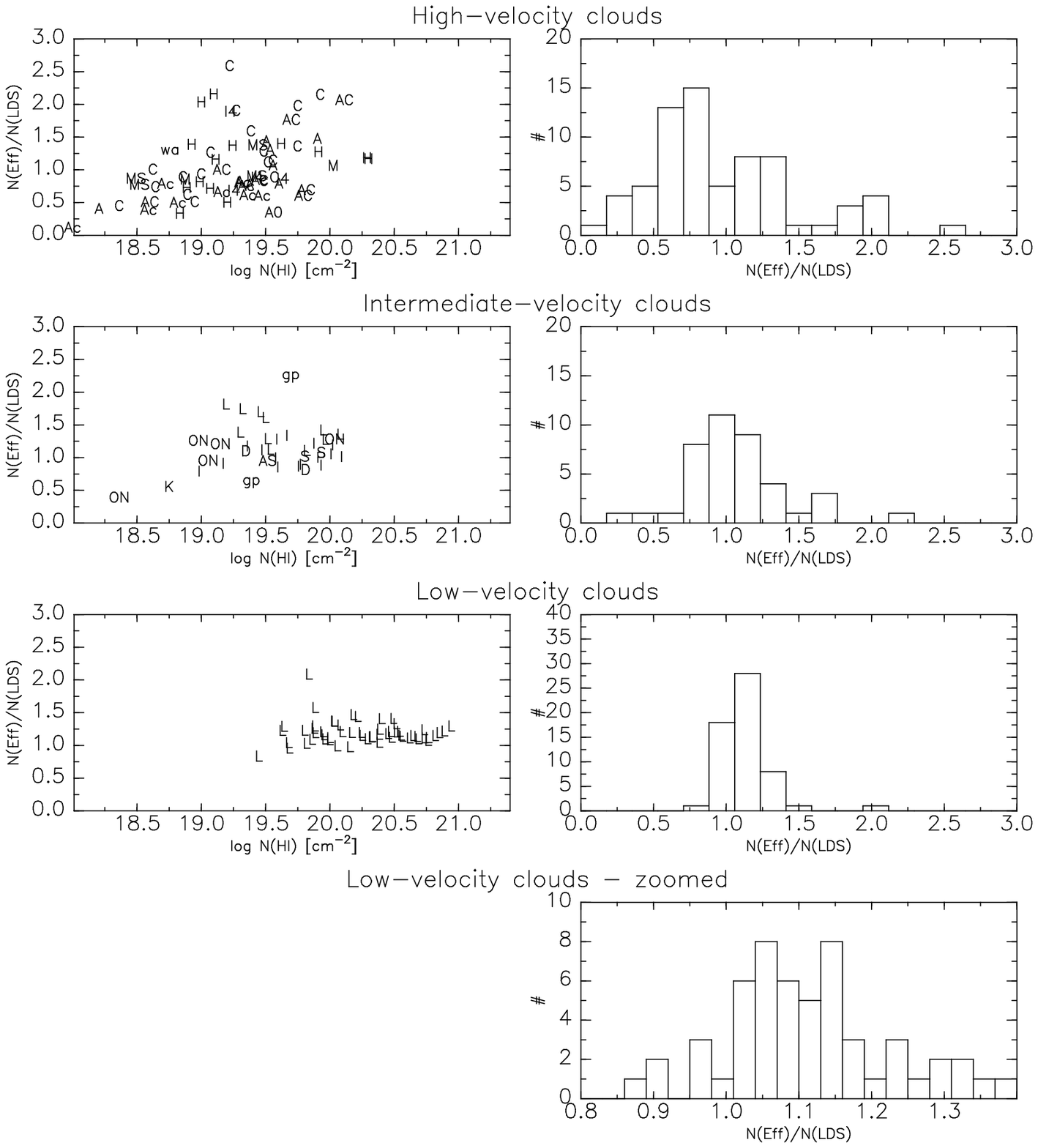}{-45}{-45}{% %AASPP
\fgnumber{3} Distribution and histogram of the ratio of column density observed
at Effelsberg and in the Leiden-Dwingeloo Survey. Top row: for high-velocity
clouds; second row: for intermediate-velocity clouds; third and bottom row: for
low-velocity gas. Left column: ratio N(\HI-eff)/N(\HI-LDS) vs N(\HI-eff). Right
column: histogram of the ratios. The letters in the plots indicate the HVC/IVC
toward which the ratio of column densities was measured.
}

\InsertPage{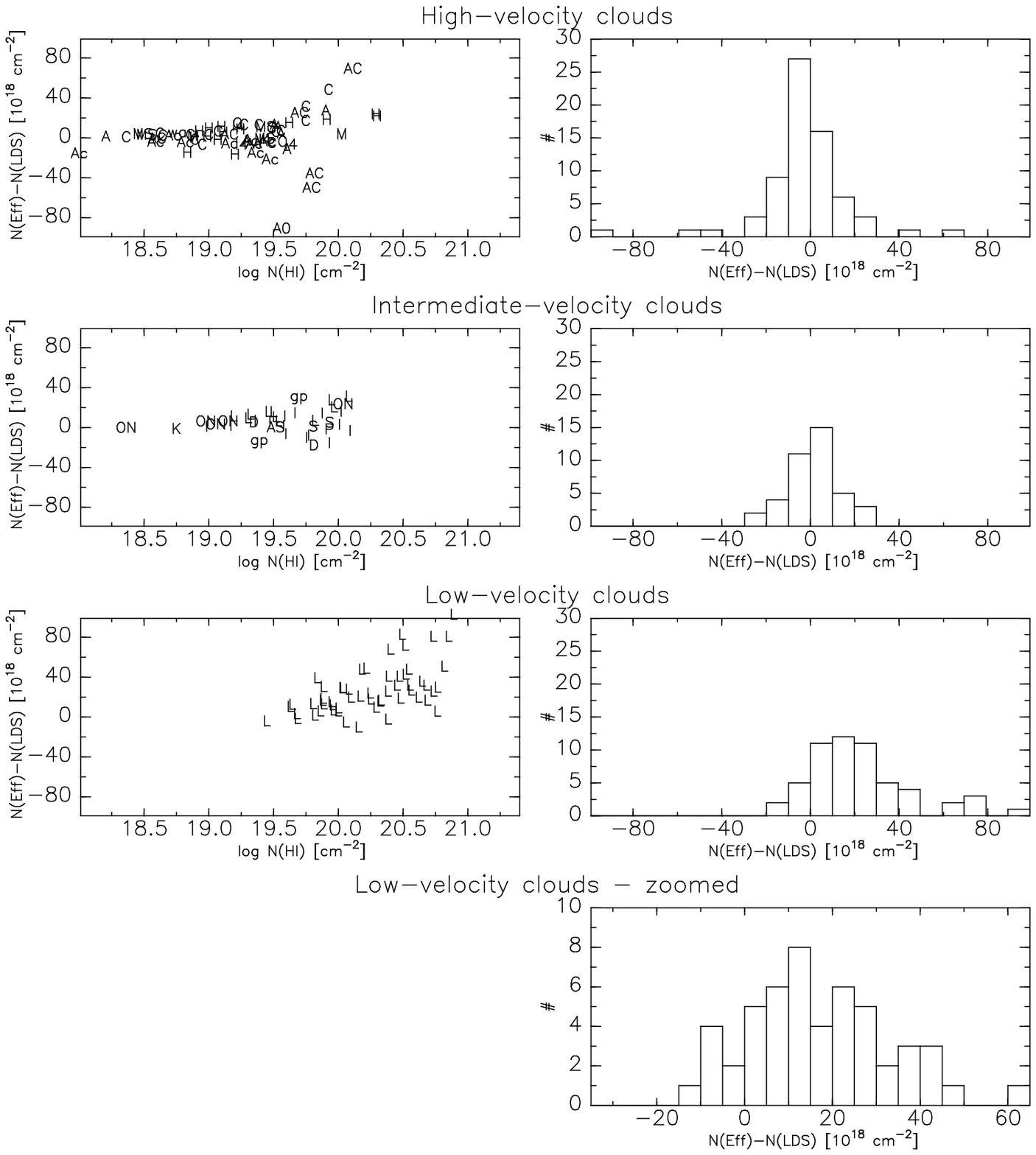}{-45}{-45}{% %AASPP
\fgnumber{4} Distribution and histogram of the difference in column density
observed at Effelsberg and in the Leiden-Dwingeloo Survey. Top row: for
high-velocity clouds; second row: for intermediate-velocity clouds; third and
bottom row: for low-velocity gas. Left column: difference
N(\HI-eff)$-$N(\HI-LDS) vs N(\HI-eff). Right column: histogram of the
differences. The letters in the plots indicate the HVC/IVC toward which the
ratio of column densities was measured.
}


\newpage
\begin{references}
\def\ref#1{\reference{DUMMY}{#1}}
\parskip=0pt
\ref {Albert C.E., Blades J.C., Morton D.C., Lockman F.J., Proulx M., Ferrarese
      L., 1993, \apjsupp, 88, 81}
\ref {Barnes D., et al. 2001, MNRAS, in press}
\ref {Braun R., Burton W.B., 2000, \aap, 354, 853}
\ref {Br\"uns C., Kerp J., Staveley-Smith L., 2001, in ``Mapping the Hidden
      Universe'', ASP Conf. Ser., in press}
\ref {Centuri\'on M., Vladilo G., de Boer K.S., Herbstmeier U., Schwarz U.J.,
      1994, \aap, 292, 261}
\ref {Colgan S.W.J., Salpeter E.E., Terzian Y., 1990, \apj, 351, 503}
\ref {Danly L., Lockman F.J., Meade M.R., Savage B.D., 1992, \apjsupp, 81, 125}
\ref {de Boer K.S., Rodriguez-Pascual P., Wamsteker W., Sonneborn G., Fransson C.,
      Bomans D.J., Kirshner R.P., 1993, \aap, 280, L15}
\ref {Diplas A., Savage B.D., 1994, \apjsupp, 93, 211}
\ref {Hartmann D., Burton W.B., 1997, Atlas of Galactic Neutral Hydrogen,
      Cambridge University Press}
\ref {Haynes R., Staveley-Smith L., Mebold U., Kalberla P.M.W., Jones K., White
      G., Jones P., Filipovic M., Dickey J., Green A., 1999, IAU Symp. 190, 108}
\ref {Kalberla P.M.W., Mebold U., Reich W., 1980, \aap, 82, 175}
\ref {Kalberla P.M.W., Mebold U., Reif K., 1982, \aap, 106, 190}
\ref {Kuntz K.D., Danly L., 1996, \apj, 457, 703}
\ref {Lilienthal D., Meyerdierks H., de Boer K.S., 1990, \aap, 240, 487}
\ref {Lockman F.J., Savage B.D., 1995, ApJS, 97, 1}
\ref {Meyer D.M., Roth K.C., 1991, \apj, 383, L41}
\ref {Payne H.E., Dickey J.M., Salpeter E.E., Terzian Y., 1978, \apj, 221, L95}
\ref {Payne H.E., Salpeter E.E., Terzian Y., 1980, \apj, 240, 499}
\ref {Putman M.E., et al., 2001, in preparation}
\ref {Ryans R.S.I., Sembach K.R., Keenan F.P., 1996, \aap, 314, 609}
\ref {Ryans R.S.I., Keenan F.P., Sembach K.R., Davies R.D., 1997a, \mnras, 289, 83}
\ref {Ryans R.S.I., Keenan F.P., Sembach K.R., Davies R.D., 1997a, \mnras, 289, 986}
\ref {Savage B.D., Wakker B.P., Bahcall J.N., Bergeron J., Boksenberg A., Hartig
      G.F., Jannuzi B.T., Kirhakos S., Lockman F.J., Murphy E.M., Sargent W.L.W.,
      Schneider D.P., Turnshek D., Weymann R.J., Wolfe A.M., 2000, \apjsupp,
      129, 563}
\ref {Schwarz U.J., Oort J.H., 1981, \aap, 101, 305}
\ref {Schwarz U.J., Wakker B.P., van Woerden H., 1995, \aap, 302, 364}
\ref {Schwarz U.J., Wakker B.P., 2001, in ``High-Velocity Clouds'', Kluwer
      Acad.\ Publ., eds.\ H.\ van Woerden, U.J.\ Schwarz, K.S.\ de Boer, B.P.\
      Wakker, in preparation}
\ref {Shaw C.R., Bates B., Kemp S.N., Keenan F.P., Davies R.D., Roger R.S., 1996,
      \apj, 473, 849}
\ref {Staveley Smith L., 1997, Proc. Ast. Soc. Austr., 14, 111}
\ref {Stoppelenburg P.S., Schwarz U.J., van Woerden H., 1998, \aap, 338, 200}
\ref {Tamanaha C.M., 1996, \apjsupp, 104, 81}
\ref {Wakker B.P., Howk C., van Woerden H., Schwarz U.J., Beers T.C., Wilhelm R.,
      Kalberla P.M.W., Danly L., 1996, \apj, 473, 834}
\ref {Wakker B.P., Howk C., Savage B.D., Tufte S.L. Reynolds R.J., van Woerden
      H., Schwarz U.J., Peletier R.F., Kalberla P.M.W., 1999b, Nature, 400, 388}
\ref {Wakker B.P., Schwarz U.J., 1991, \aap, 250, 484}
\ref {Wakker B.P., 2001, Paper~I, this Volume}
\ref {Wakker B.P., Oosterloo T.A., Putman M.E., 2001, submitted to AJ}
\ref {Wakker B.P., van Woerden H., 1997, \araa, 35, 217}
\end{references}
\end{document}